\documentclass[iop,numberedappendix]{emulateapj}

\newcommand{\degree}{$^{\circ}$}
\newcommand{\rascend}[4]{#1$^{\mathrm{h}}$#2$^{\mathrm{m}}$#3$^{\mathrm{s}}$.#4}
\newcommand{\decline}[4]{#1\degree~#2$\arcmin$~#3~#4$\arcsec$}
\newcommand{\hi}{H~{\sc i}}

\shorttitle{Diffuse X-rays in Compact Groups}
\shortauthors{Desjardins et al.}

\begin{document}

\title{Some Like It Hot: Linking Diffuse X-ray Luminosity, Baryonic Mass, and Star Formation Rate in Compact Groups of Galaxies}

\author{Tyler~D.~Desjardins\altaffilmark{1}, Sarah~C.~Gallagher\altaffilmark{1}, Ann~E.~Hornschemeier\altaffilmark{2}, John~S.~Mulchaey\altaffilmark{3}, Lisa~May~Walker\altaffilmark{4}, William~N.~Brandt\altaffilmark{5}, Jane~C.~Charlton\altaffilmark{5}, Kelsey~E.~Johnson\altaffilmark{4}, Panayiotis Tzanavaris\altaffilmark{2,6}}
\affil{}
\altaffiltext{1}{Department of Physics and Astronomy, University of Western Ontario, London, ON N6A 3K7, Canada}
\altaffiltext{2}{Laboratory for X-ray Astrophysics, NASA/Goddard Space Flight Center, Greenbelt, MD 20771, USA}
\altaffiltext{3}{Carnegie Observatories, 813 Santa Barbara Street, Pasadena, CA 91101, USA}
\altaffiltext{4}{Department of Astronomy, University of Virginia, Charlottesville, VA 22904, USA}
\altaffiltext{5}{Department of Astronomy and Astrophysics, The Pennsylvania State University, University Park, PA 16802, USA}
\altaffiltext{6}{Department of Physics and Astronomy, The Johns Hopkins University, Baltimore, MD 21218, USA}

\begin{abstract}
We present an analysis of the diffuse X-ray emission in 19 compact groups of galaxies (CGs) observed with  {\em Chandra}. The hottest, most X-ray luminous CGs agree well with the galaxy cluster X-ray scaling relations in $L_X-T$ and $L_X-\sigma$, even in CGs where the hot gas is associated with only the brightest galaxy. Using {\em Spitzer} photometry, we compute stellar masses and classify HCGs~19, 22, 40, and 42 and RSCGs 32, 44, and 86 as fossil groups using a new definition for fossil systems that includes a broader range of masses. We find that CGs with total stellar and \hi\ masses $\gtrsim10^{11.3}$~M$_\odot$ are often X-ray luminous, while lower-mass CGs only sometimes exhibit faint, localized X-ray emission. Additionally, we compare the diffuse X-ray luminosity against both the total UV and 24~\micron\ star formation rates of each CG and optical colors of the most massive galaxy in each of the CGs. The most X-ray luminous CGs have the lowest star formation rates, likely because there is no cold gas available for star formation, either because the majority of the baryons in these CGs are in stars or the X-ray halo, or due to gas stripping from the galaxies in CGs with hot halos. Finally, the optical colors that trace recent star formation histories of the most massive group galaxies do not correlate with the X-ray luminosities of the CGs, indicating that perhaps the current state of the X-ray halos is independent of the recent history of stellar mass assembly in the most massive galaxies.
\end{abstract}


\section{Introduction}
In the local Universe, the majority of galaxies exist in gravitationally bound systems, i.e., in groups or clusters (e.g.,~\citealt{tully87,small99,karachentsev05}). Cosmological $\Lambda$CDM models would imply that in rich groups and clusters of galaxies the fraction of baryons in stars may be as little as 20\% \citep{borgani04}, while observations indicate a value closer to 10\% (e.g.,~\citealt{balogh01,lin03}). The remaining baryons are in the form of gas in various states, i.e., molecular, neutral, or ionized. In rich clusters, ram-pressure stripping and harassment of gas-rich galaxies deposits vast quantities of neutral gas into the intracluster medium (ICM), providing material for a virialized X-ray halo (cf.,~\citealt{gunn72}).

In galaxy clusters, the hot ICM is already largely developed, therefore we must look to the building blocks of clusters to examine the early stages of the growth of the hot gas halos. Compact groups (CGs) have high galaxy number densities similar to the cores of rich clusters, and they are expected to experience enhanced tidal encounters and mergers compared to loose groups while their low velocity dispersions lengthen the timescales over which these encounters occur relative to clusters. These systems provide excellent laboratories to study the effects of galaxy interactions on the build-up of hot gas halos in low-mass groups of galaxies, which are the building blocks of rich clusters \citep{peebles70,gonzalez05}.

Several studies have investigated the X-ray properties of CGs and specifically the diffuse, hot gas in these systems (e.g.,~\citealt{helsdon01,desjardins13,fuse13}). The first comprehensive examination of diffuse X-ray emission in CGs was performed by \citet{ponman96}, in which the authors use X-ray observations with the {\em ROSAT} Position Sensitive Proportional Counter (PSPC) to examine the group-linked hot gas in a sample of 85 Hickson CGs (HCGs; \citealt{hickson82}) of which 22 were detected. The authors made efforts to mask the soft X-ray emission from the individual galaxies and report that the remaining emission appears to be clumpy, suggesting that, in contrast to clusters, the hot gas is not in equilibrium. 

With the much improved spatial and spectral resolution of the {\em Chandra} Advanced CCD Imaging Spectrograph (ACIS) compared to the {\em ROSAT} PSPC (angular resolutions 0\farcs5 and 25\arcsec\ FWHM, respectively), \citet{desjardins13} find that the detectable diffuse X-ray emission in a small sample of nine HCGs have varied morphologies that range broadly from linked to the individual galaxies to a true intragroup medium (IGM; not to be confused with the intergalactic medium). The galaxy-linked emission is typically associated with vigorous star formation, while hot gas in the form of an IGM is likely due to virialization of the baryons by the group potential well. HCG~42 may be an exception as it has a hot gas halo associated with the brightest group galaxy resembling a hot IGM that appears small in extent, but which may extend farther than is detectable due to low surface brightness.

In this study, we expand upon the analysis presented by \citet{desjardins13} using {\em Chandra} ACIS archival observations of an additional 10 CGs, thus bringing the total {\em Chandra} sample to 19 CGs when combined with \citet{desjardins13}. We also utilize {\em Spitzer}, Sloan Digital Sky Survey (SDSS), and Apache Point Observatory (APO) data to characterize how the total CG stellar mass relates to the observed X-ray properties of the groups. Section~\ref{sec:twosample} describes the sample selection and the general characteristics of the CGs in our study. In Section~\ref{sec:twoanalysis}, we list the {\em Chandra} data reduction steps and determine group stellar masses. Section~\ref{sec:twodiscuss} discusses our findings and their implications, and Section~\ref{sec:twoconclude} summarizes our conclusions. Errors are reported at the 90\% confidence level unless otherwise stated. For all calculations, we assumed the currently favored cosmological parameters of $\Omega_{\mathrm{M}}=0.27$, $\Omega_\Lambda=0.73$, and $H_0=70$~km~s$^{-1}$~Mpc$^{-1}$ \citep{hinshaw13}.

\section{Sample Description}
\label{sec:twosample}

\begin{deluxetable*}{lccccccccl}
\tablecolumns{10}
\tablecaption{Compact Group Sample\label{tab:sample}}
\tablehead{\colhead{Group} & \multicolumn{2}{c}{Coordinates (J2000)} & \colhead{$z$} & \colhead{v$_{\text{CMB}}$\tablenotemark{a}} & \colhead{$\sigma$\tablenotemark{b}} & \colhead{$\tilde{R}$\tablenotemark{c}} & \colhead{M$_{\text{H~{\sc i}}}$} & \colhead{M$_{\text{dyn}}$} &\colhead{References}\\
\colhead{} & \colhead{$\alpha$} & \colhead{$\delta$} & \colhead{} & \colhead{(km s$^{-1}$)} & \colhead{(km s$^{-1}$)} & \colhead{(kpc)} & \colhead{($10^{9}$ M$_\odot$)} & \colhead{($10^{11}$ M$_\odot$)} & \colhead{}}
\startdata
HCG 30 & 04$^{\text{h}}$33$^{\text{m}}$28$^{\text{s}}$ & $-02^\circ$49\arcmin57\arcsec & 0.0154 & 4562 & $121^{+24}_{-23}$ & $81.2\pm2.2$ & $0.60\pm0.06$ & $13.80\pm0.37$ & 1, 3, 4, 7\\
HCG 37 & 09$^{\text{h}}$13$^{\text{m}}$35$^{\text{s}}$ & $+30^\circ$00\arcmin51\arcsec & 0.0223 & 6940 & $451\pm17$ & $42.4\pm2.8$ & $5.40\pm0.15$ & $100.25^{+6.63}_{-6.61}$ & 1, 3, 4, 8, 9\\
HCG 40 & 09$^{\text{h}}$38$^{\text{m}}$54$^{\text{s}}$ & $-04^\circ$51\arcmin07\arcsec & 0.0223 & 7026 & $148\pm15$ & $37.6\pm0.8$ & $6.60\pm0.15$ & $9.63\pm0.21$ & 1, 3, 4, 10, 11\\
HCG 51 & 11$^{\text{h}}$22$^{\text{m}}$21$^{\text{s}}$ & $+24^\circ$17\arcmin35\arcsec & 0.0258 & 8051 & $268^{+22}_{-23}$ & $93.7\pm3.2$ & $<4.57$ & $78.39\pm2.68$ & 1, 3, 5, 8, 12\\
HCG 68 & 13$^{\text{h}}$53$^{\text{m}}$41$^{\text{s}}$ & $+40^\circ$19\arcmin07\arcsec & 0.0080 & 2583 & $108^{+8}_{-9}$ & $60.0\pm2.2$ & $5.60\pm0.02$ & $7.21\pm0.30$ & 1, 2, 3, 4, 13, 14, 15, 16\\
HCG 79 & 15$^{\text{h}}$59$^{\text{m}}$12$^{\text{s}}$ & $+20^\circ$45\arcmin31\arcsec & 0.0145 & 4439 & $265^{+11}_{-12}$ & $16.5\pm1.0$ & $4.20\pm0.13$ & $13.41\pm0.81$ & 1, 2, 3, 4, 10, 17, 18\\
HCG 97 & 23$^{\text{h}}$47$^{\text{m}}$23$^{\text{s}}$ & $-02^\circ$19\arcmin34\arcsec & 0.0218 & 6174 & $359\pm18$ & $99.8\pm5.9$ & $4.3\pm0.13$ & $149.63^{+8.87}_{-8.86}$ & 1, 2, 3, 4, 9, 11, 19\\
HCG 100 & 00$^{\text{h}}$01$^{\text{m}}$21$^{\text{s}}$ & $+13^\circ$07\arcmin57\arcsec & 0.0178 & 4976 & $142\pm24$ & $54.1\pm1.5$ & $9.0\pm0.09$ & $12.65\pm0.36$ & 1, 3, 4, 20, 21\\
RSCG 17 & 01$^{\text{h}}$56$^{\text{m}}$22$^{\text{s}}$ & $+05^\circ$38\arcmin37\arcsec & 0.0190 & 5415 & $386\pm19$ & $35.3\pm2.4$ & $<15.85$ & $61.14^{+4.23}_{-4.21}$ & 2, 6, 9, 12\\
RSCG 31 & 09$^{\text{h}}$17$^{\text{m}}$22$^{\text{s}}$ & $+41^\circ$57\arcmin24\arcsec & 0.0060 & 2009 & $52\pm12$ & $41.3\pm1.1$ & 1.82 & $1.28\pm0.04$ & 2, 6, 22, 23, 24
\enddata
\tablenotetext{a}{Value taken from the NASA Extragalactic Database (NED) based upon \citet{fixsen96}.}\\
\tablenotetext{b}{We recalculated the velocity dispersion $\sigma$ using the CG member velocities from the cited references.}
\tablenotetext{c}{The median projected two-galaxy separation.}
\tablerefs{Catalogs: (1)~\citet{hickson82}; (2)~\citet{barton96} --- Positions: (3)~\citet{hickson92} --- \hi\ Masses: (4)~\citet{borthakur10}; (5)~\citet{verdes01}; (6) L.~M.~Walker~et. al. (in preparation) --- Velocities: (7)~\citet{jones09}; (8)~\citet{falco99}; (9)~\citet{mahdavi04}; (10)~\citet{nishiura00}; (11)~\citet{decarvalho97}; (12)~\citet{rc3}; (13)~\citet{vandriel01}; (14)~\citet{denicolo05}; (15)~\citet{rhee96}; (16)~\citet{simien02}; (17)~\citet{paturel03}; (18)~\citet{bofanti99}; (19)~\citet{mahdavi05}; (20)~\citet{huchra99}; (21)~\citet{hickson93}; (22)~\citet{strauss92}; (23)~\citet{monnier03}; (24)~\citet{nordgren97}}
\end{deluxetable*}

The selection criteria of HCGs and Redshift Survey CGs (RSCGs; \citealt{barton96}) yield samples that are well suited for our study of the hot gas properties in dense galaxy groups. The HCG catalog was compiled by searching Palomar Observatory Sky Survey (POSS) data, which cover the sky at all declinations $\delta>-27^\circ$, for groups with $N\ge4$ galaxies\footnote{This requirement has been relaxed to $N\geq3$ due to the discovery as a result of spectroscopic follow-up that only 69\% of HCGs have $N\geq4$ accordant members, while 92\% have $N\geq3$ \citep{hickson92}.} with magnitudes within 3~mag of the brightest group galaxy, $\theta_N\geq3\theta_G$, and $\bar{\mu}_G<26$~mag~arcsec$^{-2}$. In this definition, $\theta_N$ is the angular diameter of the largest concentric circle that does not include non-member galaxies within 3~mag of the brightest group galaxy, $\theta_G$ is the angular diameter of the smallest concentric circle that includes the nuclei of all of the group members, and $\bar{\mu}_G$ is the surface brightness averaged over the circle defined by $\theta_G$. The last two criteria for the HCG catalog pertain to the isolation and compactness of the groups, respectively. All photometric measurements were made in the POSS $E$-band (most equivalent to the standard Johnson $R$ filter), for which the POSS observations are complete to $m=20.0$~mag.

\citet{barton96} used the second Center for Astrophysics redshift survey (CfA2) and second Southern Sky Redshift Survey (SSRS2) to identify a sample of CGs with similar properties to the \citet{hickson82} catalog. The CfA2 consists of a strip that covers approximately $117^\circ\times6^\circ$ and is centered near the north Galactic pole (e.g.,~\citealt{delapparent86}), while the SSRS2 covers 1.3~steradians around the southern Galactic cap \citep{dacosta94}. Both surveys are complete to $m_{B_0}=15.5$ and \citet{barton96} only consider galaxies in the line-of-sight velocity range $300\leq v\leq 15000$~km~s$^{-1}$. These RSCGs are selected using a friends-of-friends algorithm in which groups with $N\geq3$ are considered RSCGs, and group members are found using $V_0\leq1000$~km~s$^{-1}$ and $D_0\leq50$~kpc, where $V_0$ and $D_0$ represent the velocity difference and projected separation, respectively, between neighbor galaxies. \citet{barton96} chose the value of $D_0$ to most closely match the observed properties of the HCG sample.

Our sample consists of all HCGs and RSCGs available in the {\em Chandra} data archive that are not part of our previous diffuse X-ray study presented in \citet{desjardins13} and that are completely covered by the {\em Chandra} footprint. We then refine our sample by comparing RSCGs with nearby galaxy clusters and removing those groups with projected separations of $<1$~Mpc and velocity differences $<3\sigma$ from a galaxy cluster. The group mean position and velocity were used for comparison. This is necessary because we wish to include only {\em isolated} CGs, and the RSCG catalog does not employ an isolation criterion similar to that of the HCGs. Due to the lack of such a criterion, the RSCG catalog includes a number of  dense environments misidentified as CGs, e.g.,~RSCGs~67 and 68 are in the core of the Coma Cluster. The criteria described above resulted in a sample of 10 additional CGs in our study compared to \citet{desjardins13}, which we list in Table~\ref{tab:sample}. We note, however, that \citet{desjardins13} did include HCG~42, which is a high galaxy density region potentially located in a filamentary structure along the line-of-sight \citep{konstantopoulos13}. Indeed, \citet{diazgimenez10} suggest that at best $\sim85$\% of HCGs with more than four members are truly dense systems rather than chance alignments. This leads to an estimate that 2--3 of the groups in our sample may not be spatially dense, however some of our systems only have three galaxies, which complicates this estimate. Further, our sample is not completely random as 16 of the groups between our study and \citet{desjardins13} were selected to study X-ray emission in the group environment (with HCGs~51 and 97 and RSCG~17 chosen because they were known to be X-ray bright), while the remaining groups were observed to study either X-ray emission in and around early-type galaxies or the supernova SN2006jc in RSCG~31. 

\begin{deluxetable*}{lcccccc}
\tablecolumns{7}
\tablewidth{0pt}
\tablecaption{Extended Group Membership\label{tab:extended}}
\tablehead{\colhead{Group} & \colhead{$r_{500}\tablenotemark{a}$} & \colhead{Galaxy} & \colhead{$m_{\mathrm{R}}$} & \colhead{Ang. Sep.} & \colhead{Reference} & \colhead{Dwarfs} \\
\colhead{} & \colhead{(arcmin)} & \colhead{} & \colhead{(mag)} & \colhead{(arcmin)} & \colhead{} & \colhead{}}
\startdata
HCG~30 & 11.0 & $\cdots$ & $\cdots$ & $\cdots$ & $\cdots$ & 0 \\
HCG~31 & 12.5 & $\cdots$ & $\cdots$ & $\cdots$ & $\cdots$ & 2\tablenotemark{b}\\
HCG~37 & 11.7 & $\cdots$ & $\cdots$ & $\cdots$ & $\cdots$ & 0 \\
HCG~40 & 7.4 & $\cdots$ & $\cdots$ & $\cdots$ & $\cdots$ & 2\tablenotemark{c}\\
HCG~51 & 9.3 & $\cdots$ & $\cdots$ & $\cdots$ & $\cdots$ & 7\\
HCG~68 & 13.6 & $\cdots$ & $\cdots$ & $\cdots$ & $\cdots$ & 2\\
HCG~79 & 11.3 & $\cdots$ & $\cdots$ & $\cdots$ & $\cdots$ & 1\\
HCG~92 & 14.5 & $\cdots$ & $\cdots$ & $\cdots$ & $\cdots$ & 3\\
HCG~100 & 10.1 & Mrk~935 & 14.28 & 1.8 & 1 & 0\\
RSCG~31 & 25.0 & $\cdots$ & $\cdots$ & $\cdots$ & $\cdots$ & 1
\enddata
\tablenotetext{a}{Or the angular separation corresponding to a projected 200~kpc radius, whichever is larger.}
\tablenotetext{b}{One of the two dwarf galaxies in HCG~31 is actually low-mass tidal debris designated HCG~31R \citep{verdes05}.}
\tablenotetext{c}{One of the two dwarf galaxies in HCG~40 (HCG~40-06 in \citealt{decarvalho97}) has only a $B$-band magnitude listed in NED, corresponding to an absolute magnitude $M_B\approx-17$~mag \citep{decarvalho97}, but no $R$-band data is listed. Based on an inspection of the image, and the fact that the $J$-, $H$-, and $K$-band magnitudes from the 2MASS catalog \citep{twomasscat} are all $>3$~mag fainter than HCG~40A, we classify this as a dwarf group member.}
\tablerefs{(1)~\citet{hickson89b}}
\end{deluxetable*}

Another concern when studying CGs is the possibility of additional galaxies far from the compact core, but still bound to the group. \citet{decarvalho97}, \citet{zabludoff98}, and \citet{konstantopoulos10,konstantopoulos12,konstantopoulos13} examined the extended populations of HCGs~7, 16, 22, 40, 42, 59, 62, 90, and 97 and RSCG~17 and found that only HCGs~42, 62, and 90 and RSCG~17 had substantial populations of galaxies outside of the core region, while HCG~97 is missing two relatively bright galaxies in the HCG catalog. We therefore label these groups as lower-limits with respect to their total group stellar masses in the figures throughout this work. For the remaining groups in our sample, we used the NASA Extragalactic Database (NED) to search for additional galaxies within the larger of $r_{500}$ or 200~kpc in radius and $\pm1000$~km~s$^{-1}$ of the group mean velocity. The results of this search are show in Table~\ref{tab:extended}. Note that we only provide detailed information on the luminous galaxies, and simply list the aggregrate number of dwarfs in each group. To distinguish between luminous and dwarf group members, we applied an absolute magnitude cut at $-17$~mag in the $R$-band. We further required that the galaxy be within three magnitudes of the brightest group galaxy to satisfy the \citet{hickson82} selection criteria, as any galaxies failing this test would likely be of little relative importance in determining total group properties. Finally, in some cases, photometric data were missing in NED, and we examined the images by eye to compare the relative sizes of the galaxies on the sky (this resulted in only dwarf galaxy classifications). Only HCG~100 excludes a relatively massive galaxy in the Hickson catalog, while HCG~51 is missing 7 dwarf galaxies. We label these two additional CGs as lower-limits in total group stellar mass in our figures. Note that for CGs embedded within larger structures, we only consider the properties of galaxies that make up the compact region. While this exclusion of the extended populations may seem in error, \citet{palumbo95} examined the extended populations of the \citet{hickson82} sample and found that the compact cores and extended halos showed statistically different properties (e.g.,~spiral fraction) indicating that the compact groups are ``disconnected'' from their environments. Evidence of this distinction between CG galaxies and their surrounding environment can be seen in the work of \citet{johnson07}, \citet{walker10}, and \citet{walker12} who found a gap in the mid-IR color distribution of CG galaxies suggestive of accelerated evolution attributed to the CG environment. Further, the galaxies far from the compact cores are, in most cases, dwarf galaxies that do not add significant stellar mass to the group. Dozens of such galaxies would be required to significantly affect our results. While the group members far outside the core may also add substantially to the total group star formation rate, these members are not yet impacted by ram-pressure stripping nor have they contributed much gas to the formation of the intragroup medium, therefore we exclude them in the discussion of the link between star formation and diffuse X-ray luminosity.

We also include a comparison sample of galaxy clusters from \citet{wu99} and \citet{zhang11}. We selected clusters from \citet{wu99} and \citet{zhang11} for each of the $L_X-T$, $L_X-\sigma$, and $\sigma-T$ relationships by including only clusters that had published uncertainties for both values in each scaling relation (for further details see \citealt{desjardins13}). The \citet{wu99} clusters are amassed from the literature (see their Table~1 for the full list of references), and have redshifts $z<1$ and $\langle z\rangle\approx0.1$, temperatures $1\lesssim T\lesssim17$~keV, velocity dispersions $150\lesssim \sigma\lesssim 2000$~km~s$^{-1}$, and X-ray luminosities $42\lesssim\log_{10}(L_X)\lesssim46$. The \citet{zhang11} measurements use {\em XMM-Newton} observations of 62 of the 64 HIghest X-ray FLUx Galaxy Cluster Sample (``HIFLUGCS''; \citealt{reiprich02}) galaxy clusters, which were originally identified using {\em ROSAT} X-ray data at Galactic latitudes of $|\ell|>20^\circ$. The clusters from \citet{zhang11} have luminosities $42\lesssim\log_{10}(L_X)\lesssim45$~erg~s$^{-1}$, temperatures $0.7\lesssim T\lesssim15$~keV, velocity dispersions $200\lesssim\sigma\lesssim1000$~km~s$^{-1}$, and a mean redshift of $\langle z\rangle=0.05$. The X-ray properties of the \citet{zhang11} clusters are measured within $r_{500}$, while the \citet{wu99} clusters are taken from the literature and corrected to a common radius using a $\beta$ model. Though it is unclear what radius \citet{wu99} used, it is reasonable to assume this correction was performed to $r_{500}$. 

\section{Analysis}
\label{sec:twoanalysis}
\subsection{X-ray Observations}

The {\em Chandra} observations are summarized in Table~\ref{tab:obs}. Data were taken in either FAINT or VFAINT mode with no gratings. We performed the calibration of the {\em Chandra} data using the {\em Chandra} Interactive Analysis of Observations (CIAO version 4.5) tool in conjunction with the CIAO calibration database version 4.5.5.1. Beginning with the Level~1 events file, we processed the data using {\tt acis\_process\_events} with corrections for the charge transfer inefficiency and time-dependent gain. We used the status bits in the Level~1 events file set by the standard data processing pipeline tasks {\tt destreak} and {\tt acis\_find\_afterglow}. The pixel randomization normally used in the {\em Chandra} data pipeline was removed to prevent degradation of the spatial resolution. A 0\farcs5 pixel randomization is necessary for data with exposure times of $\lesssim2$~ks to compensate for aliasing effects, however the exposure times of the observations in our sample are typically far in excess of this limit, therefore we omitted the randomization and recover the resolution to subtract robustly the point sources from the diffuse emission.

For data taken in VFAINT mode, we applied the VFAINT cleaning algorithm, which uses a $5\times5$ pixel event island, rather than the $3\times3$ island in FAINT mode, for the rejection of cosmic rays. The VFAINT cleaning method has been shown to occasionally reject photons from bonafide X-ray point sources leading to underestimates of the associated fluxes, however this does not impact our analysis of the diffuse emission. We then filtered the data on the standard {\em Advanced Satellite for Cosmology and Astrophysics} ({\em ASCA}) grades and selected only grades 0, 2, 3, 4, and 6 to produce the final Level~2 events file for analysis.

\begin{deluxetable*}{lcccccl}
\tablecolumns{7}
\tablecaption{Summary of {\em Chandra} ACIS Observations\label{tab:obs}}
\tablehead{\colhead{Group} & \colhead{ObsID} & \colhead{Array} & \colhead{Mode} & \colhead{Exposure} & \colhead{Date} & \colhead{Previous}\\
\colhead{} & \colhead{} & \colhead{} & \colhead{} & \colhead{(ks)} & \colhead{} &\colhead{Publications}}
\startdata
HCG 30 & \dataset[ADS/Sa.CXO#obs/06977]{6977} & S &VF & 29.7 & 2006-02-07 & 1, 2\\
HCG 37 & \dataset[ADS/Sa.CXO#obs/05789]{5789} & S &VF & 17.9 & 2005-01-13 & 1, 2\\
HCG 40 & \dataset[ADS/Sa.CXO#obs/05788]{5788} & S & VF & 33.2 & 2005-01-29 & 1, 2, 3\\
	      & \dataset[ADS/Sa.CXO#obs/06203]{6203} & S & VF & 15.0 & 2005-01-29 & 1, 3\\
HCG 51 & \dataset[ADS/Sa.CXO#obs/04989]{4989} & S & VF & 38.5 & 2004-02-15 & 2, 4--10\\
	      & \dataset[ADS/Sa.CXO#obs/05304]{5304} & S& VF & 13.0 & 2005-02-24 & 5, 6, 9\\
HCG 68 & \dataset[ADS/Sa.CXO#obs/05903]{5903} & S& VF & 4.5 & 2005-04-10 & 3, 11\\
HCG 79 & \dataset[ADS/Sa.CXO#obs/11261]{11261} & S & VF & 69.2 & 2010-05-20 & 12\\
HCG 97 & \dataset[ADS/Sa.CXO#obs/04988]{4988} & S & VF & 57.4 & 2005-01-14 & 1, 2, 9, 13\\
HCG 100 & \dataset[ADS/Sa.CXO#obs/06978]{6978} & I &VF & 27.8 & 2006-12-06 & 2\\
	        & \dataset[ADS/Sa.CXO#obs/08491]{8491} & I &VF & 17.8 & 2007-01-24 & 1\\ 
RSCG 17 & \dataset[ADS/Sa.CXO#obs/02223]{2223} & S& F & 30.4 & 2001-01-28 & 4, 5, 7, 9, 10, 14--28\\
RSCG 31 & \dataset[ADS/Sa.CXO#obs/06729]{6729} & S& VF & 54.5 & 2007-01-06 & 29, 30\\
	         & \dataset[ADS/Sa.CXO#obs/08457]{8457} & S& F & 9.8 & 2006-11-04 & 11, 29--31\\
	         & \dataset[ADS/Sa.CXO#obs/09093]{9093} & S &VF & 24.8 & 2008-01-20 & 30\\
	         & \dataset[ADS/Sa.CXO#obs/10567]{10567} & S& F & 5.1 & 2009-01-24 & 30--32
\enddata
\tablerefs{(1)~\citet{rasmussen08}; (2)~\citet{fuse13}; (3)~\citet{cluver13}; (4)~\citet{kim07a}; (5)~\citet{sun09a}; (6)~\citet{sun09b}; (7)~\citet{dong10}; (8)~\citet{haggard10}; (9)~\citet{sun12}; (10)~\citet{trichas12}; (11)~\citet{liu11}; (12)~\citet{tamburri12}; (13)~\citet{eckmiller11}; (14)~\citet{diehl05}; (15)~\citet{fukazawa06}; (16)~\citet{humphrey06}; (17)~\citet{jetha07}; (18)~\citet{kim07b}; (19)~\citet{diehl07}; (20)~\citet{rasmussen07}; (21)~\citet{jetha08}; (22)~\citet{jeltema08}; (23)~\citet{diehl08}; (24)~\citet{wang10}; (25)~\citet{giacintucci11}; (26)~\citet{matsushita12}; (27)~\citet{crain13}; (28)~\citet{heida13}; (29)~\citet{gibson12}; (30)~\citet{ofek13}; (31)~\citet{immler06}; (32)~\citet{grier11}; (33)~\citet{smith12}}
\end{deluxetable*}

Before examining the diffuse emission, we first excised the point sources from the observations. We used the Mexican hat wavelet detection routine {\tt wavdetect} \citep{freeman02} to search for point sources in the data. Multiple {\em Chandra} obsIDs were merged using {\tt merge\_obs} prior to running the detection algorithm to facilitate the detection of faint point sources. The merged images were created after correcting the aspect solution to compensate for small offsets in the World Coordinate System between the multiple observations.

To detect robustly all of the point sources in the data, we followed a prescription similar to that presented in \citet{tullmann11}. Specifically, we divided the data into the energy ranges 0.5--2 (``soft''), 2--8 (``hard''), and 0.5--8~keV (``full'') with block~1, 2, 4, and 8~pixel spatial binning in each energy range. This produced 12 images on which to run {\tt wavdetect}. We generated a point-spread function (PSF) model for each position on the CCDs of interest using the {\tt mkpsf} routine in CIAO with an encircled energy fraction of 95\% at the midpoint of each energy range. The source significance threshold was dynamically set such that there was approximately one false source detected per wavelet scale in each image. Specifically, we used the {\em falsesrc} parameter in {\tt wavdetect} to allow the source significance threshold for each pixel to vary. However, we note that individual, unbinned pixels cannot be used for source detection as {\tt wavdetect} suppresses fluctuations on scales smaller than the PSF. We chose wavelet scales for source detection of $2^{n/2}$ for integers $n$ such that $0\leq n\leq3$. Our goal for the point source detection was to be extremely conservative and reject all potential point sources because, if present in the extracted spectra of the diffuse emission, they introduce strong biases in the results. We then matched the resulting point source catalogs using an angular separation tolerance of 0\farcs5, first selecting the smallest spatial binning scale in which the source was detected in an energy band, ensuring the best centroid position for the source. These three catalogs were then matched with the same tolerance across the energy bands choosing the sources with the smallest PSF. This creates one point source catalog per compact group.

Extraction regions for the CGs were selected to include all of the member galaxies as well as any obvious diffuse X-ray emission. As discussed in \citet{desjardins13}, the emission in most CGs is clearly not virialized, and therefore an extraction region defined according to the virial radius (e.g.,~$r_{500}$) is not physically meaningful in these systems. Larger extraction regions simply result in additional noise and lead to larger uncertainties in our subsequent spectral-model fitting. Our extraction regions are defined for each group in Table~\ref{tab:regions}.

As in \citet{desjardins13}, we used the ACIS stowed background data for determination of the instrumental background. We note that in groups with low signal-to-noise, excess residual line emission was observed in the background-subtracted spectra at $\sim1.8$~keV. A strong line at this energy is observed in the stowed background data, and we attribute the excess emission in the science spectra to under-subtraction of  an instrumental feature, e.g., the Si~K line at 1.845~keV or the iridium edge in the 1.8--2.1~keV range.

\begin{deluxetable}{lrrcc}
\tablecolumns{5}
\tablecaption{Extraction Region Parameters\label{tab:regions}}
\tablewidth{0pt}
\tablehead{\colhead{Group} & \multicolumn{2}{c}{J2000 Coordinates} & \colhead{Shape} & \colhead{Radius}\\
\colhead{} & \colhead{$\alpha$} & \colhead{$\delta$} & \colhead{} & \colhead{}}
\startdata
HCG~30 & \rascend{4}{36}{25}{9} & \decline{--2}{50}{14}{5} & Circular & 4\farcm4\\
HCG~37 & \rascend{9}{13}{36}{2} & \decline{29}{59}{24}{3} & Circular & 3\farcm3\\
HCG~40 & \rascend{9}{38}{55}{2} & \decline{--4}{51}{2}{1} & Circular & 3\farcm8\\
HCG~51 & \rascend{11}{22}{21}{8} & \decline{24}{17}{39}{5} & Circular & 3\farcm5\\
HCG~68 & \rascend{13}{53}{36}{7} & \decline{40}{18}{52}{6} & Circular & 6\farcm3\\
HCG~79 & \rascend{15}{59}{11}{5} & \decline{20}{45}{26}{2} & Circular & 4\farcm1\\
HCG~97 & \rascend{23}{47}{25}{6} & \decline{--2}{19}{5}{6} & Elliptical & 3\farcm7$\times$3\farcm0\tablenotemark{a}\\
HCG~100 & \rascend{0}{1}{20}{0} & \decline{13}{7}{2}{8} & Circular & 3\farcm5\\
RSCG~17 & \rascend{1}{56}{21}{6} & \decline{5}{37}{53}{6} & Circular & 3\farcm7\\
RSCG~31 & \rascend{9}{17}{23}{4} & \decline{41}{57}{17}{7} & Circular & 4\farcm3
\enddata
\tablenotetext{a}{The position angle is $0^\circ$.}
\end{deluxetable}

Events were extracted using the \anchor{http://www.astro.psu.edu/xray/acis/acis\_analysis.html}{{\tt ACIS Extract}} ({\tt AE}) software package\footnote{The {\em ACIS Extract} software package and User's Guide are available at \url{http://www.astro.psu.edu/xray/acis/acis\_analysis.html}.}\citep{broos10,broos12}. The point source catalogs for each group were input into {\tt AE} for PSF modeling using {\tt MARX} version 4.4. The point sources were then excised from the events files prior to extracting the CG spectra. Specifically, we used {\tt AE} to create a circular mask for each point source that enclosed 99\% of the PSF, and then multiplied the mask radius by a factor of 1.1 to ensure no contamination of the diffuse emission by the wings of the PSF. In addition, the point source masks were also applied to the stowed background data for the extraction of the background spectra. We used {\tt AE} to generate response matrix files and the {\em CIAO} tool {\tt mkwarf} to create weighted ancillary response files using the weight map extension of the spectral files. 

\begin{deluxetable*}{lccccccc}
\tablecolumns{8}
\tablecaption{Best-Fitting X-ray Model Parameters\label{tab:xraypars}}
\tablehead{& {\tt tbabs} & {\tt ztbabs} & \multicolumn{3}{c}{{\tt MEKAL}} & & \\
\colhead{Group} & \colhead{Galactic H~{\sc i}} & \colhead{Redshifted H~{\sc i}} & \colhead{$kT$} & \colhead{$Z/Z_{\odot}$} & \colhead{A\tablenotemark{a}} & \colhead{$L_X$\tablenotemark{b}} & \colhead{$\chi^2$/d.o.f.}\\
\colhead{} & \colhead{($10^{20}$ cm$^{-2}$)} & \colhead{($10^{20}$ cm$^{-2}$)} & \colhead{(keV)} & \colhead{} & \colhead{($10^{-4}$~cm$^{-5}$)} & \colhead{(erg s$^{-1}$)} & \colhead{}}
\startdata
\object{HCG 30} & 4.72 & $\cdots$ & 0.6 & 0.5 & $<0.47$ & $<40.59$ & $\cdots$ \\
\object{HCG 37} & 1.87 & $\cdots$ &$1.02^{+0.14}_{-0.11}$ & $1.21^{+>1}_{->1}$ & $0.75^{+0.80}_{-0.75}$ & $41.36^{+0.31}_{-2.91}$ & 98.28/72\\
\object{HCG 40} & 3.60 & $\cdots$ & 0.6 & 0.5 & $<0.30$ & $<40.76$ & $\cdots$ \\
\object{HCG 51} & 1.13 & $10.10^{+1.74}_{-1.69}$ & $1.36\pm0.04$ & $0.29\pm0.04$ & $29.06^{+1.94}_{-1.83}$ & $42.73\pm0.03$ & 100.97/77\\
\object{HCG 68} & 0.96 & $\cdots$ & $0.57^{+0.07}_{-0.08}$ & $14.58^{+>1}_{->1}$ & $2.36^{+3.24}_{3.05}$& $41.13\pm0.06$ & 52.32/32\\
\object{HCG 79} & 3.86 & $\cdots$ & 0.6 & 0.5 & $<0.28$ & $<40.54$ & $\cdots$\\
\object{HCG 97} & 3.61 & $18.57\pm3.14$ & $0.85\pm0.03$ & $0.19\pm0.02$ & $28.72^{+2.85}_{-2.58}$ & $42.45\pm0.04$ & 311.61/145\\
\object{HCG 100} & 4.51 & $\cdots$ & 0.6 & 0.5 & $<0.47$ & $<40.66$ & $\cdots$ \\
\object{RSCG 17} & 4.33 & $4.93^{+1.31}_{-1.25}$ & $1.15^{+0.04}_{-0.03}$ & 0.11 & $21.78^{+2.54}_{-2.23}$ & $42.22\pm0.02$ & 207.47/138\\
\object{RSCG 31} & 1.15 & $\cdots$ & 0.6 & 0.5 & $<0.24$ & $<39.70$ & $\cdots$
\enddata
\tablenotetext{\ }{{\bf Notes.} Data for additional CGs can be found in \citet{desjardins13}.}
\tablenotetext{a}{The normalization of the MEKAL model.}
\tablenotetext{b}{The X-ray luminosity over the range 0.01--100~keV.}
\end{deluxetable*}

Extracted spectra were then fit in {\tt XSPEC} version 12.7.1 using a combination of foreground absorption and a thermal plasma. Prior to fitting, we binned the spectra using the {\tt HEASoft} tool {\tt grppha} such that each bin had a minimum of 20 counts; this ensures that Gaussian statistics (i.e.,~$\chi^2$ fitting) may be used. The multiplicative Tuebingen-Boulder interstellar medium absorption model ({\tt tbabs}) was used to account for photoelectric absorption along the line of sight. For this purpose, we used the relative abundances from \citet{lodders03}. The Galactic hydrogen column density was fixed to the value from the weighted average of the \citet{kalberla05} \hi\ maps using the {\tt HEASoft} {\tt nH} tool. The thermal plasma was modeled using the {\tt MEKAL} plasma model \citep{mewe85,mewe86,kaastra92,kaastra93,liedahl95,kaastra95} with the ionization balance from \citet{arnaud85} and \citet{arnaud92}. The low energy resolution non-grating spectra are insufficient to determine plasma densities, therefore we fixed the density in the model to a reasonable value of $n=1$~cm$^{-3}$. We calculated the X-ray luminosities over the range 0.01--100~keV using a ``dummy'' response created by the {\tt XSPEC} command {\tt dummyrsp}. The results of the best-fitting spectral models are reported in Table~\ref{tab:xraypars}. 

In HCGs~51 and 97, and RSCG~17, the model overestimated the X-ray emission below $\sim$0.7~keV. We used HCG~97 to test three different additional model components: (1) a second {\tt MEKAL} plasma; (2) a simple power law; and (3) additional absorption at the redshift of the groups ({\tt ztbabs} in {\tt XSPEC}). The second plasma component of case (1) failed to fit the observed flux at low energies; however, the power law and the additional absorption produced nearly equal values of $\chi_\nu^2=2.0$ and 2.1, respectively. In both cases (2) and (3), the temperatures were identical within the errors. In case (2), the power law had a hard photon index of 1.4, which led to a higher luminosity compared to the additional absorption model ($\log_{10}[L_{X,pl}]-\log_{10}[L_{X,ztbabs}]=0.41$) due to increased flux at higher energies. We chose to use the additional absorption component and find redshifted \hi\ column densities in HCGs~51 and 97, and RSCG 17 of $1.01\times10^{21}$, $1.86\times10^{21}$, and $4.93\times10^{20}$~cm$^{-2}$, respectively, corresponding to \hi\ masses of $\sim10^5$~M$_\odot$. We note that HCGs~51 and 97 have very extended X-ray emission (see Section~\ref{subsec:twomorph}), and if the absorption interpretation is correct, we may be detecting low-surface brightness cool gas on the near sides of these systems. Indeed, this would be consistent with the \hi\ upper-limit of HCG~97 with Very Large Array $L$-band imaging (S. Borthakur, private communication). The X-ray emission in RSCG~17 subtends a much smaller angle compared to HCG~97, but may still be explained with the low-surface brightness interpretation. Comparatively, the inclusion of a hard power law component does not have a physical motivation, but cannot be summarily ruled out because of the low signal at $E\gtrsim3$~keV.

If the number of X-ray photons associated with the CG was less than $3\sigma$ above the instrumental background, we classified such a source as a non-detection and used a plasma temperature of $T=0.6$~keV to set an upper-limit on the X-ray luminosity, identical to the method presented in \citet{desjardins13}. We use the definition of $\sigma$ from \citet{desjardins13} such that $\sigma=\left[SB+(A_st_s/A_bt_b)B\right]^{1/2}$, where $SB$ is the total counts in the source before background subtraction, $B$ is the number of counts in the background, $A$ is the area of the extraction region, $t$ is the integration time, and the subscripts $s$ and $b$ represent the science and background observations, respectively.

Note that HCG~79, also known as Seyfert's Sextet, lies at Galactic coordinates $\ell=35.0^\circ$ and $b=46.9^\circ$ and is coincident with a portion of the North Polar Spur (NPS). The NPS is a region of bright, soft X-ray emission associated with expanding supernovae remnants (e.g.,~\citealt{cruddace76,borken77,iwan80,miller08}), therefore making it more difficult to detect emission from HCG~79. In addition, an X-ray bright background group or cluster of galaxies with $z\sim0.3$ is located 0\farcm6 to the northwest of HCG~79 \citep{palma02,tamburri12}. Rather than spatially model and exclude the emission from this background source, we included it in the spectral extraction and then modeled it with an additional plasma component.

\subsection{Optical Data}

To compare the diffuse X-ray and the relative, optical brightnesses of the two brightest group galaxies in each CG (see Section~\ref{sec:baryons}), we used observations of HCGs~16, 19, 26, 33, 40, 42, 48, and 62, and RSCG~15, obtained on 18 January 2011 at the APO 3.5-meter telescope using the Seaver Prototype Imaging camera (SPIcam) instrument and the SDSS $r^\prime$ filter. SPIcam is a $2048\times2048$ pixel CCD with a scale of 0\farcs14/pixel; however, because the APO site is seeing-limited, we used $2\times2$ pixel binning to facilitate faster readout time without sacrificing spatial resolution. For the remaining CGs in the \citet{walker12} Expanded Sample, we used $r^\prime$-band images taken from the SDSS DR9 database. We include all of the CGs in the Expanded Sample rather than only the CGs observed with {\em Chandra} to have a statistically large enough sample to compare against the X-ray groups. Note that between the APO and SDSS observations, we have optical coverage of the entire Expanded Sample except HCGs~90 and 91, and RSCG~4. This left us with stellar mass measurements for galaxies in 47 CGs. The optical photometry of CGs using SDSS data is further explored in \citet{walker13}.

We reduced the APO $r^\prime$-band data using {\tt PyRAF} version 2.0 and {\tt IRAF}\footnote{{\tt IRAF} is distributed by the National Optical Astronomy Observatory, which is operated by the Association of Universities for Research in Astronomy (AURA) under cooperative agreement with the National Science Foundation.} version 2.14 to perform serial overscan subtraction and to create master two-dimensional bias and dark frames, as well as master flat images in both filters. We used Source Extractor \citep{sextractor} version 2.8.6 for all extended source photometry with an aperture set to twice the Petrosian radius to ensure uniformity across the sample. No absolute photometric calibration was performed as we were only interested in the differential photometry of the two brightest group galaxies. The small projected separations of galaxies in CGs necessitated that we be particularly careful with object blending. In cases where multiple sources overlap, Source Extractor uses a de-blending algorithm to separate the pixels associated with each object. We found that the default de-blending parameters were sufficient for E/S0 galaxies, however we needed to adjust the settings on an individual basis for inclined, star-forming galaxies to ensure that the entire galaxy was classified as one source rather than a collection of blended sources.

Combined with the SDSS DR9 images, we have optical photometry in the $r^\prime$-band for 41 CGs, all but one of which are in the \citet{walker12} Expanded Sample, while HCG~51 is solely in our X-ray sample. Note that HCG~30, which is in our X-ray sample, but not the \citet{walker12} Expanded Sample, does not have APO or SDSS data. We use the \citet{hickson82} ordering of CG galaxies according to their optical brightness to select the first and second rank galaxies with respect to the POSS $E$-band luminosity. For the RSCGs, we use the ordering presented by \citet{walker13}, which uses the same ordering system albeit in a marginally different bandpass compared to the HCGs.

\subsection{Stellar Mass Determination}
\label{subsec:stellarmass}

\begin{figure}[t]
\begin{center}
\includegraphics[width=\columnwidth]{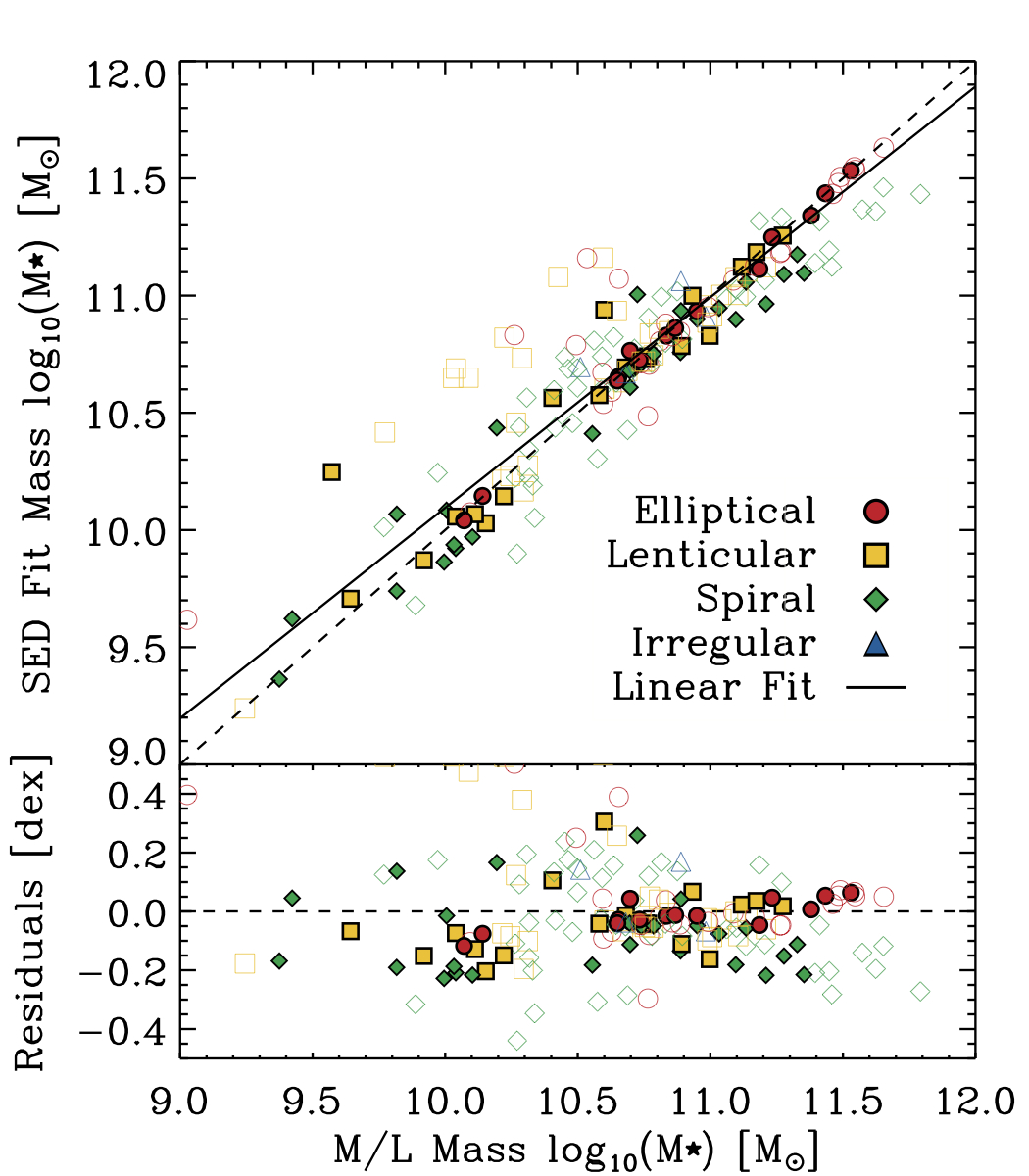}
\caption{Comparison of the stellar masses determined by mid-IR SED fitting against the masses using the $K$-band M/L relationship from \citet{bell03}. Galaxies in this X-ray study are plotted with filled symbols while the remaining galaxies from the Expanded Sample defined by \citet{walker12} are plotted as open symbols for comparison. The dashed line in the upper panel shows the one-to-one relation for reference, while the solid line is the orthogonal distance regression fit to the data. The bottom panel shows the residuals around the fit with a dashed line for reference. Most of the observed scatter is likely due to scatter in the M/L relation, as well as uncertainties in the SED fitting.\label{fig:masscompare}}
\end{center}
\end{figure}

The standard method of stellar mass determination, i.e., use of the $K_s$-band luminosity, assumes a universal mass-to-light ratio independent of the galaxy morphology and, therefore, star formation history. To determine more robustly the stellar masses of the galaxies in our sample, we used the library of galaxy templates generated by the {\tt GRASIL}\footnote{\url{http://adlibitum.oat.ts.astro.it/silva/grasil/grasil.html}} code \citep{silva98,silva99,granato00,bressan02,silva03,panuzzo03,vega05,silva09} to fit the galaxy spectral energy distributions (SEDs). Two Micron All Sky Survey (2MASS) $JHK_s$ and {\em Spitzer} Infrared Array Camera (IRAC) 3.6--8.0~\micron\ fluxes were taken from \citet{walker12}. In the case of saturation in one of the four IRAC bands (6 galaxies), or if IRAC data were missing (9 galaxies), we calculated the stellar mass using only the $K_s$-band M/L relation from \citet{bell03}, i.e., M$_\odot$/L$_{\nu,\odot}=0.95\pm0.03$. For galaxies missing 2MASS data from \citet{walker12}, we used the 2MASS photometry from the {\em WISE} database. For all sources, we converted the fluxes to luminosities using distances determined from the 3~K cosmic microwave background (CMB) dipole-corrected velocities. All of the elliptical and spiral templates, as well as the M82 starburst galaxy template, were fit to the data without knowing {\em a priori} the galaxy morphology to ensure unbiased results. We note that the best-fitting templates do agree well with the observed galaxy morphologies in a general sense, i.e., spiral galaxies are best modeled using spiral templates and likewise for elliptical galaxies, though the finer divisions within these classes (e.g., Sa, Sb, Sc) are sometimes not accurately determined from the SED fitting.

In the near-infrared, the SED of a galaxy scales with stellar mass, therefore the normalization of the best fit galaxy template to the observed luminosity coupled with the stellar mass of the model yields the stellar mass of the galaxy. To properly fit the templates to the galaxy photometry, we first shifted the {\tt GRASIL} SED templates to the observed frame of the source and then convolved them with the 2MASS and {\em Spitzer} filter response curves. We used the normalization to the 3.6~\micron\ luminosity as an initial guess of the stellar mass normalization before $\chi^2$ minimization. We note that mid-IR active galaxies contain strong polycyclic aromatic hydrocarbon emission features in the 5.8 and 8.0~\micron\ IRAC bands, however these did not greatly affect the quality of the SED fitting. We estimated the errors on the SED-fitted masses by varying the normalization of the template until $\Delta\chi^2=2.71$, i.e., the 90\% confidence interval. The total stellar masses for the CGs in \citet{desjardins13} and this paper are listed in Table~\ref{tab:stellar_mass}. The masses of the individual galaxies in the Expanded Sample are shown in Figure~\ref{fig:masscompare} where we compare the SED fitted masses against the stellar masses derived from the $K_s$-band M/L relationship from \citet{bell03}. We find that the stellar masses from SED fitting match well with some scatter compared to those from the $K_s$-band method, though there is a small deviation at low masses.

\begin{deluxetable}{lrc}
\tablecolumns{3}
\tablecaption{Total Group Stellar Masses Using Core Galaxies\label{tab:stellar_mass}}
\tablehead{\colhead{Group} & \colhead{Stellar Mass} & \colhead{$N_{\mathrm{gal}}$}\\
\colhead{} & \colhead{($10^9$ M$_{\odot}$)} & \colhead{}}
\startdata
HCG 7    &   $334.39\pm 0.34$                      &    4\\
HCG 16    &   $462.24\pm 1.92$                       &    4\\
HCG 22    &   $104.58\pm 0.23$                      &    3\\
HCG 30    &   $0.88\pm 0.00$                      &    4\\
HCG 31    &   $32.37\pm 0.26$                      &    3\\
HCG 37    &   $460.67\pm 0.28$                      &    5\\
HCG 40    &   $398.47\pm 0.40$                      &    5\\
HCG 42    &   $458.91\pm 0.36$                      &    4\\
HCG 51    &   $49.68\pm 0.00$                       &    5\\
HCG 59    &   $36.05\pm 0.00$                       &    4\\
HCG 62    &   $229.34\pm 0.37$                      &    4\\
HCG 68    &   $427.37\pm 0.56$                      &    5\\
HCG 79    &   $79.56\pm 0.01$                     &    4\\
HCG 90    &   $308.05\pm 1.61$                       &    4\\
HCG 92    &   $523.66\pm 2.13$                       &    4\\
HCG 97    &   $328.03\pm 0.44$                      &    5\\
HCG 100    &   $127.11\pm 0.00$                       &    4\\
RSCG 17   &   $288.40\pm 2.30$                       &    3\\
RSCG 31   &   $72.96\pm 0.07$                     &    3
\enddata
\end{deluxetable}

\section{Results and Discussion}
\label{sec:twodiscuss}

We detect diffuse X-ray emission in 50\% of the CGs in our new sample observed with {\em Chandra}. Combined with the CGs from \citet{desjardins13}, this yields 19 groups with 12 detections and an overall detection rate of 63\%. We caution the reader that our detection rate should not be used to draw conclusions about the statistical distribution of diffuse X-ray luminosities of CGs (e.g.,~\citealt{ponman96}) as many of the targets that make up our sample were observed on the assumption that the groups would be X-ray bright. 

\subsection{X-ray Morphology}
\label{subsec:twomorph}

\citet{desjardins13} find that, in contrast to galaxy clusters, the diffuse X-ray emission in CGs is often linked to the individual galaxies rather than the group itself. Therefore, we construct contour maps of the X-ray emission to examine the distribution in this extended X-ray sample. To make the contour maps, we first excised the point sources and interpolated over them using the {\tt CIAO} task {\tt dmfilth}. Note that these interpolated images were only used in the creation of the contour maps and not in the spectral analysis of the diffuse X-ray emission. The resulting image of only diffuse emission was divided by the monoenergy exposure map (optimized at a photon energy of 1~keV) to create flux images in units of photons~s$^{-1}$~cm$^{-2}$. Finally, we smoothed the flux images by convolving them with a Gaussian kernel.

Figure~\ref{fig:morph} shows the diffuse X-ray contour maps for the X-ray detected CGs (see \citealt{desjardins13} for the contour maps of the CGs in that paper). We once again find a mixture of galaxy- and group-linked emission, with HCGs~51 and 97 and RSCG~17 having the most extended X-ray halos. In all cases, detected X-ray emission is centered on the optically brightest group galaxy, and is not as localized as in HCGs~16 and 31 \citep{desjardins13}. This suggests that for the X-ray detected CGs with galaxy-linked diffuse emission in this paper that were not previously presented in \citet{desjardins13}, the sources are hot gas halos around the brightest group galaxies and not star formation within the galaxies. This is further supported by the fact that all of the X-ray detected CGs in this new sample of ten groups have E/S0 brightest group galaxies.

In HCGs~68 and 97, the X-ray morphologies are indicative of recent or ongoing galaxy-galaxy interactions. In Figure~\ref{fig:morph}, one can see a hot-gas bridge connecting HCG~68A and B. Interestingly, the morphology of the X-ray gas in HCG~97 is elongated in three directions away from the brightest group galaxy toward galaxies 97D and E, and is most pronounced towards the southeast. This last direction may indicate a past tidal encounter with one of the other galaxies in the group.

\begin{figure*}
\centering
\includegraphics[width=\columnwidth]{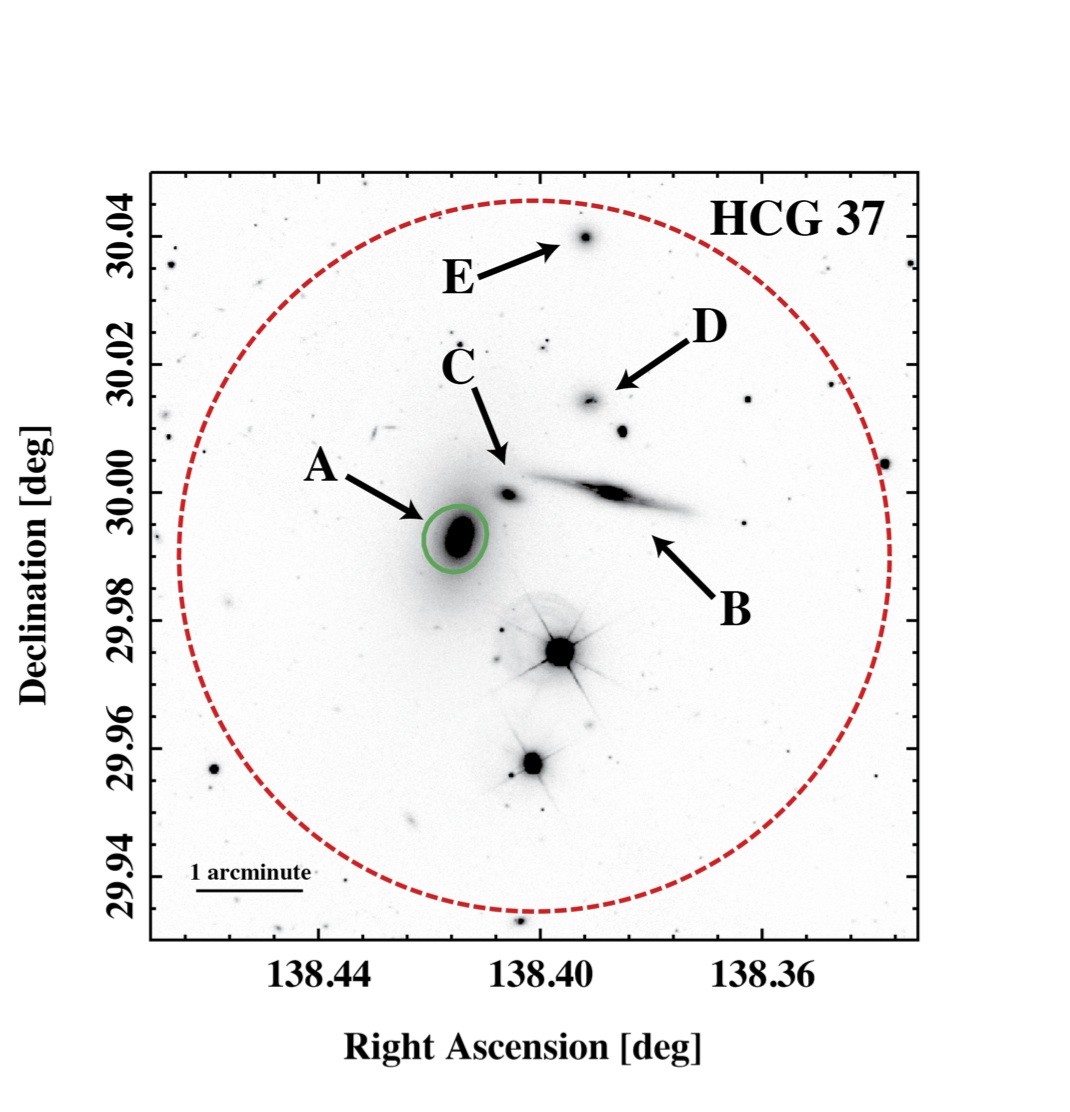}
\includegraphics[width=\columnwidth]{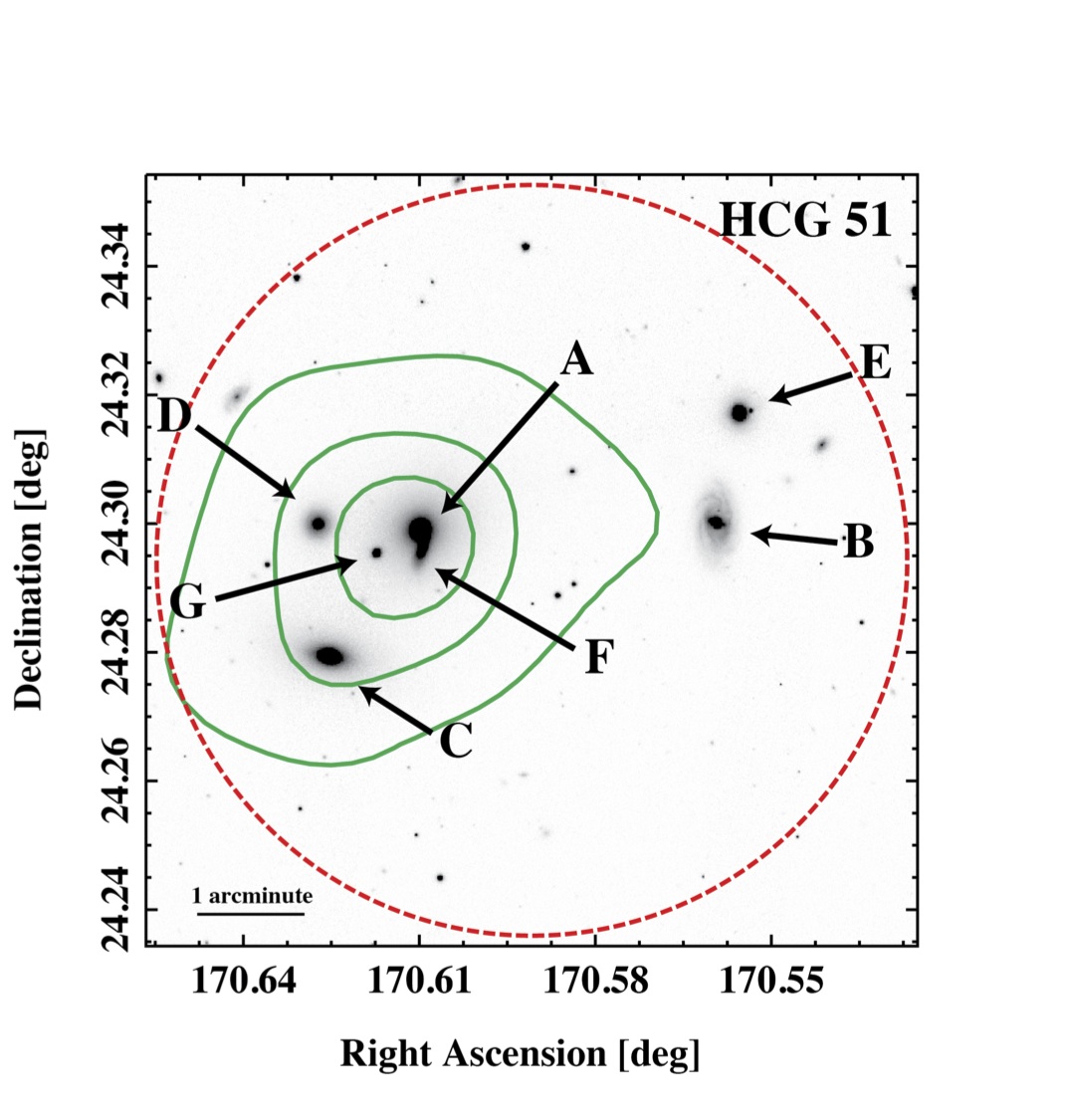}\\
\includegraphics[width=\columnwidth]{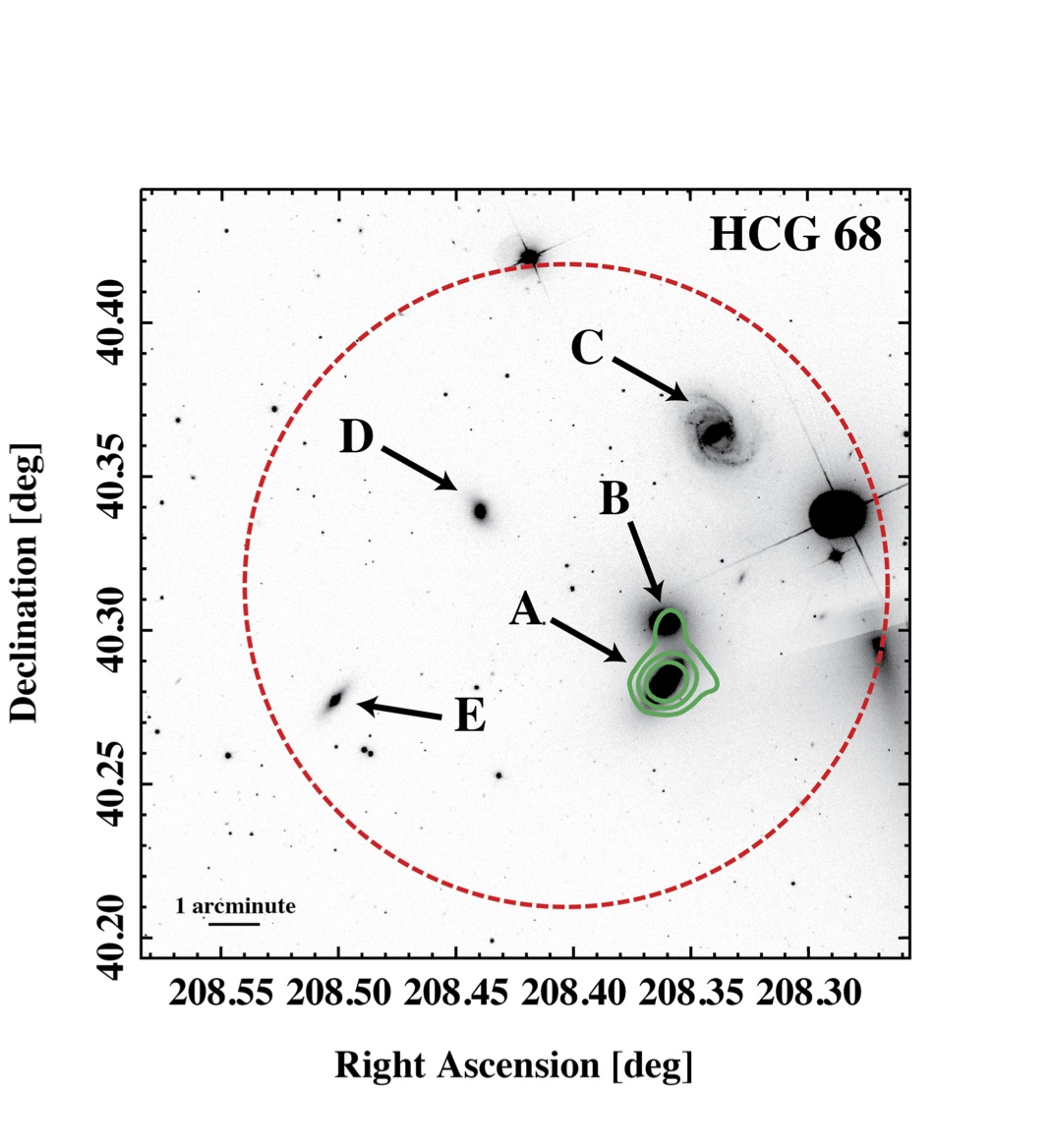}
\includegraphics[width=\columnwidth]{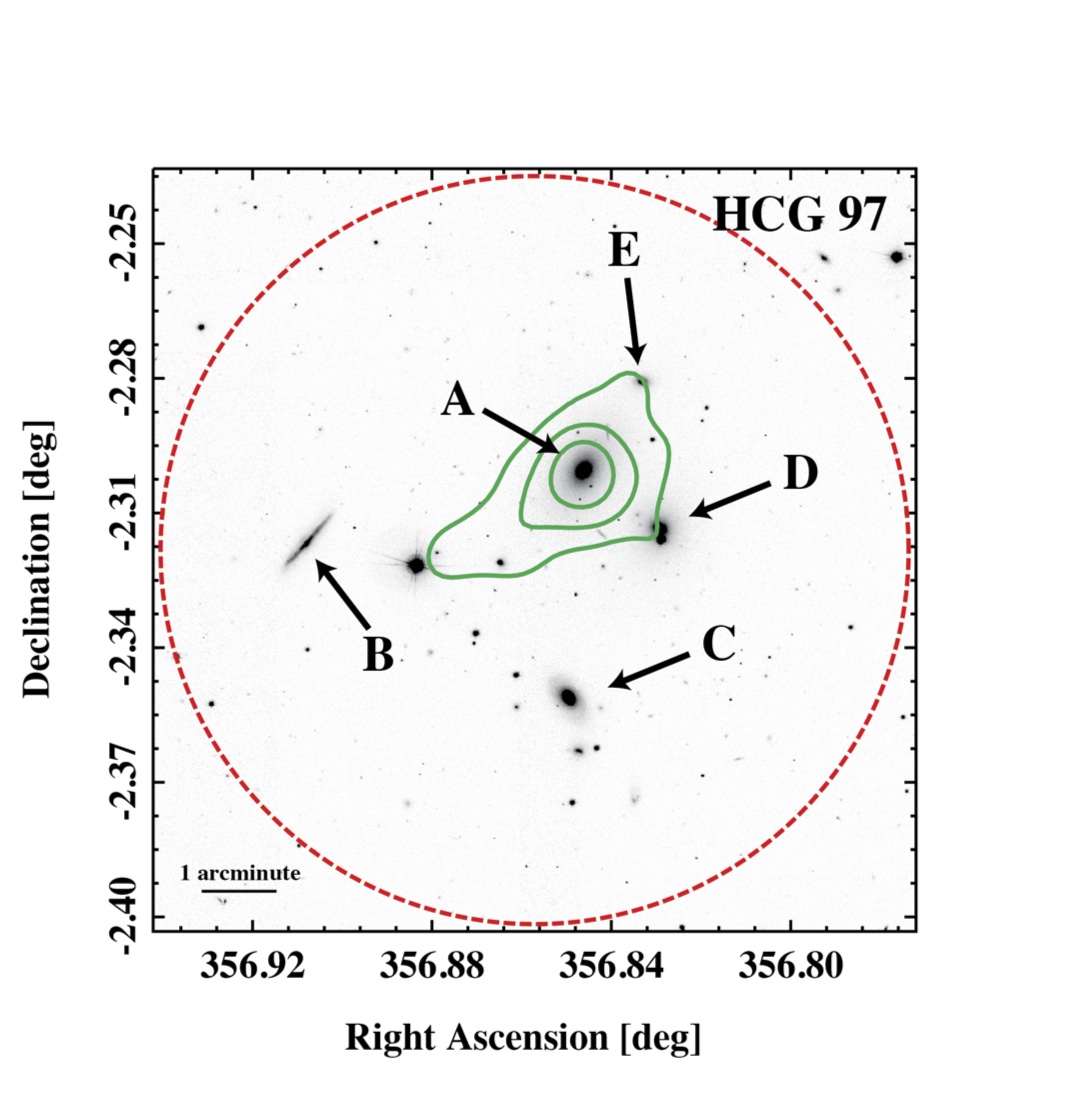}
\caption{X-ray contour maps of the CGs in the 0.5--2~keV band. The optical images come from the SDSS DR9 and are in the $r^\prime$ filter. In HCGs~37, 68, 97, and RSCG 17, the X-ray emission is marked by green contours at levels of $2.5\times10^{-8}$, $5\times10^{-8}$, $1\times10^{-7}$, $2.5\times10^{-7}$~photons~s$^{-1}$~cm$^{-2}$. In HCG~51, the green X-ray contours correspond to $1\times10^{-7}$, $1.5\times10^{-7}$, $2\times10^{-7}$, and $2.5\times10^{-7}$~photons~s$^{-1}$~cm$^{-2}$. The red dashed region indicates the outer extraction boundary for each CG. X-ray point sources are also excised, but are not labeled in these images. We label CG members with accordant redshifts using the notation from \citet{hickson82} and \citet{walker13} for reference.\label{fig:morph}}
\end{figure*}

\begin{figure*}
\addtocounter{figure}{-1}
\centering
\includegraphics[width=\columnwidth]{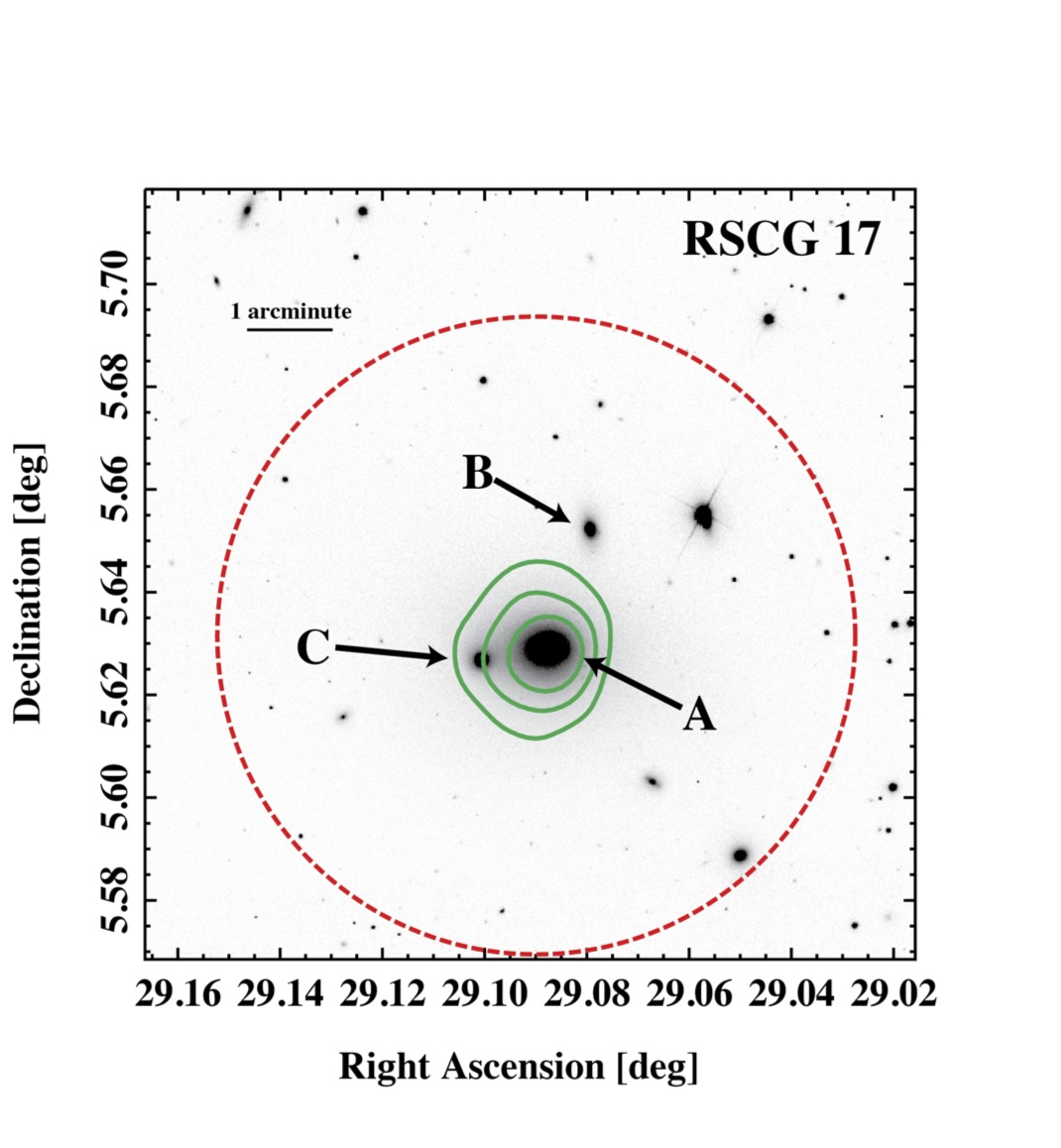}
\caption{Continued.}
\end{figure*}

\subsection{X-ray Scaling Relations}
\label{subsec:scaling}

\begin{figure*}[!th]
\centering
\includegraphics[width=\textwidth]{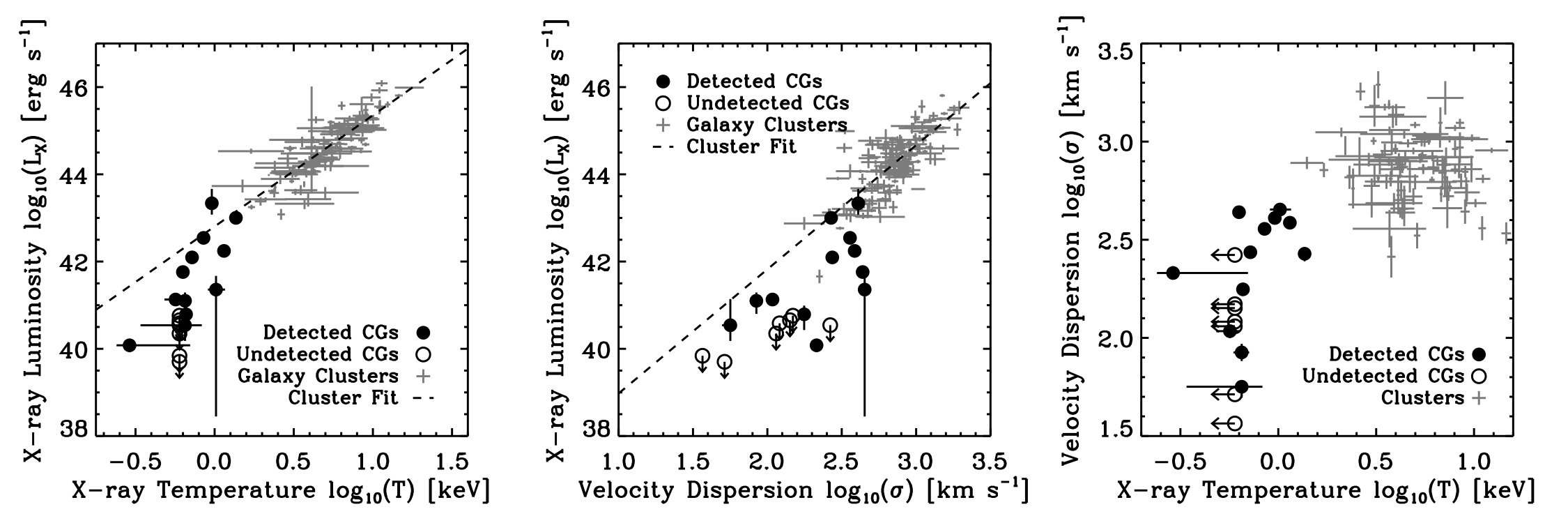}
\caption{The X-ray scaling relations for the cluster samples of \citet{wu99} and \citet{zhang11}, corrected to our cosmology, and the CGs: $L_X-T$ ({\em left}); $L_X-\sigma$ ({\em center}); and $\sigma-T$ ({\em right}). In all three panels, the gray crosses represent the galaxy clusters, while the filled and open circles are the X-ray detected and non-detected CGs, respectively, in the combined sample of this study and \citet{desjardins13}. In the $L_X-T$ and $L_X-\sigma$ diagrams, the dashed line shows the orthogonal distance regression fit to the clusters. Such a fit is not possible for the $\sigma-T$ relation because of the large scatter in the cluster data and the paucity of clusters at low velocity dispersions. For the non-detected CGs, we use a reasonable temperature of 0.6~keV for determining upper-limits in $L_X$.\label{fig:lxscale}}
\end{figure*}

We wish to investigate how the CG X-ray properties compare with the X-ray cluster scaling relations between the bolometric X-ray luminosity ($L_X$), X-ray temperature, and group velocity dispersion. This allows us to examine the physical nature of the hot plasma in groups independent of the gas morphology. These power-law scaling relations are expected from the virial theorem and the self-similar model of galaxy cluster and ICM formation \citep{kaiser86}.

After inspecting the hot gas morphologies and surface brightness profiles of the CGs, we surmise that the six most luminous CGs may have additional X-ray emission out to larger radii (cf.~HCG~62 in \citealt{desjardins13}). To compensate for this, we used a $\beta$ (hydrostatic, isothermal sphere) model to quantify the X-ray luminosity correction out to $r_{500}$. The $\beta$ model of the surface brightness profile is given as
\begin{equation}
S(R) = S_0 \left[1 + \left(\frac{R}{r_c}\right)^2\right]^{-3\beta + 0.5},
\end{equation}
where $S_0$ is the peak surface brightness, $R$ is the distance from the center, $r_c$ is the core radius (for which we have once again assumed the median two-galaxy separation). The $\beta$ term is defined as
\begin{equation}
\beta \equiv \frac{\mu m_p\sigma^2}{kT},
\end{equation}
where $\mu$ is the mean molecular weight (fixed at solar), $m_p$ is the mass of the proton, $\sigma$ is the line-of-sight velocity dispersion, $k$ is the Boltzmann constant, and $T$ is the gas temperature. This resulted in correction factors of order unity in most cases except HCG 62, which required a factor of 12.8 to compensate for the small extraction region used in \citet{desjardins13}. Note that in \citet{desjardins13}, we found a correction factor of 3.1 was necessary to scale the observed {\em Chandra} flux to that observed by ROSAT in the much larger aperture used by \citet{mulchaey98}.

The left panel of Figure~\ref{fig:lxscale} shows the CGs and the galaxy cluster sample from the literature in the $L_X-T$ plane. There exists a population of high-temperature groups ($T$$\sim$1~keV; HCGs~51, 62, 97, and RSCG~17) that agree well with the cluster $L_X-T$ relation, and two lower temperature CGs (HCGs~37 and 42) that are also in agreement. The agreement between the aforementioned hot, X-ray luminous CGs and the clusters occurs within the scatter of the cluster data. In \citet{desjardins13}, the authors find groups that agree with the cluster scaling relations are those in which the emission is linked primarily to the IGM rather than to the individual galaxies (e.g., HCG~62); however, HCG~37 represents a CG where the ionized gas is clearly associated with the brightest group galaxy, and therefore may be probing the cool-temperature, low-luminosity portion of the X-ray scaling relations. If this is true, it may indicate that HCG~42, which has a similar temperature and X-ray luminosity, is not an example of group-linked emission, but a CG in which the X-ray halo is only observed around the brightest group galaxy.

The CGs and clusters in the $L_X-\sigma$ plane are shown in the center panel of Figure~\ref{fig:lxscale}. Consistent with the $L_X-T$ relation, there exists a population of high velocity dispersion CGs which seem to agree well, albeit with more scatter, with the cluster sample from the literature, and in fact these are the same CGs as those that matched the $L_X-T$ relation. These six CGs, HCGs~37, 42, 51, 62, 97, and RSCG 17, may represent the most cluster-like CGs in our sample with respect to their hot gas properties. We again note that the consistency between the high-dispersion, X-ray luminous CGs and the clusters occurs at the level of the scatter in the cluster data, while the CGs themselves all exist systematically below the $L_X-\sigma$ relation fitted to the galaxy clusters.

For completeness, we also examine the $\sigma-T$ relationship for CGs and clusters in the right panel of Figure~\ref{fig:lxscale}, however we find that the clusters exhibit a large scatter and, due to under-sampling at lower velocity dispersions/cooler temperatures, the relationship is too poorly constrained for comparison with the CGs. We do note that there is a rapid drop in velocity dispersion at cool X-ray temperatures in the CGs, further indicating that these are not relaxed systems.

\subsection{Relating X-ray Emission to the Baryonic Mass}
\label{sec:baryons}

Understanding the relative importance of the hot, warm, and cool gas phases is critical to understanding the evolution of systems of galaxies. At a more fundamental level, the distribution and phase of the baryons in galaxy groups dictate the future evolutionary path of the system, while also giving insight into the group history. For example, a system lacking cool/cold gas while having a hot X-ray halo (e.g.,~HCG~62; \citealt{desjardins13}) is unlikely to convert much more gas into stellar mass, and therefore future galaxy evolution will be primarily dynamical. Conversely, galaxies in systems such as HCG~16 that are \hi-rich will evolve both dynamically and in terms of their stellar populations. While the reservoir of cool and cold gas in systems of galaxies is critical to these two examples, it is likely that the bulk of the baryonic mass will end up in either stars or an X-ray emitting halo\footnote{This neglects the contribution of other negligible components of baryonic mass in systems of galaxies (e.g., dust), as well as baryons that are expelled from groups.}, therefore examining the relationship between stellar mass, cool gas, and group X-ray emission gives a more complete picture of galaxy evolution in groups. We hypothesize that the mass of these systems is critical to how they will evolve, i.e., it will determine the ability of the group potential to heat gas to X-ray temperatures and consequently lower the baryon fraction in stars. An increase in the amount of hot gas between the galaxies will further affect galaxy evolution within the groups through, e.g., ram-pressure stripping.

If we disregard the dynamical masses due to their large intrinsic uncertainties (see Section~\ref{sec:xrayhi}), then we can compare the relative masses of the groups as the sum of their directly observable components. For example, HCG~22 has no detected diffuse X-ray emission ($\log_{10}[L_X]<39.84$~erg~s$^{-1}$; \citealt{desjardins13}), therefore we can assume that relatively few baryons are found in the hot IGM. Furthermore, the molecular gas mass of non-star-bursting galaxies is typically negligible compared to the combined H~{\sc i} and stellar mass (see, e.g.,~\citealt{young91}), thus the total baryonic mass of the group may be approximated as the sum of the stellar and H~{\sc i} masses (similar to \citealt{connelly12}). If we assume that the fraction of group mass in baryons, as in more massive clusters, is constant (e.g.,~\citealt{gonzalez07,andreon10}), then HCG~22 is at least a factor of 4 less massive than the X-ray luminous HCG~42. This example initially supports our hypothesis regarding the existence of an X-ray halo and its dependence on the group mass, therefore we choose to expand our test to a larger sample.

\begin{figure}[t]
\centering
\includegraphics[width=\columnwidth]{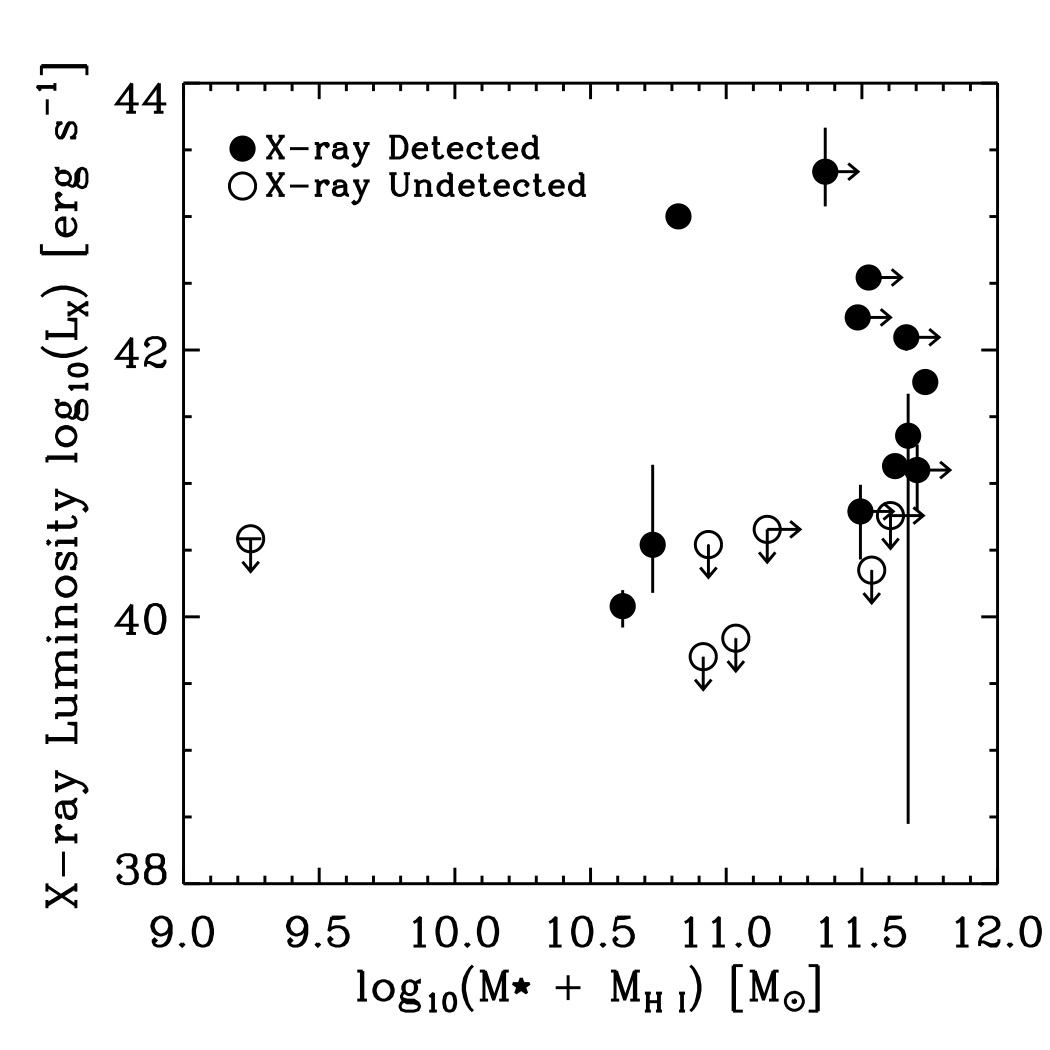}
\caption{The CG diffuse X-ray luminosity as a function of the total stellar and \hi\ mass. Assuming that the baryon fraction of groups is approximately constant, as has been seen in clusters, the total stellar and \hi\ masses then trace the total masses of low-mass systems, while they provide lower-limits on the total masses of X-ray luminous groups with larger hot gas masses. The lower-limits in the total mass arise from the exclusion of additional group members at large distances from the compact core region, but which are still within $r_{500}$.\label{fig:lx_baryon}}
\end{figure}

We plot the diffuse X-ray luminosity against the total stellar and \hi\ masses in Figure~\ref{fig:lx_baryon}. While the most X-ray luminous and cluster-like CGs likely have a substantial fraction of baryons in the hot phase, it is not possible to measure the hot gas mass of baryons without making potentially poor assumptions about the diffuse emission, e.g., spherical symmetry and the value of $\mathrm{d}T/\mathrm{d}r$, therefore we do not include the hot gas mass in our total mass estimates. Again assuming that the baryon fraction in groups is approximately constant, then the total stellar and \hi\ mass may be used as a proxy for the total group mass in X-ray faint systems, and as a lower-limit in X-ray luminous CGs. From Figure~\ref{fig:lx_baryon}, we find that nearly all of the X-ray luminous CGs with intragroup X-ray hot gas have higher total stellar and \hi\ masses than the CGs that were not detected by {\em Chandra} and those with galaxy-linked X-ray emission. The approximately vertical distribution of CGs at $M_\star+M_{\text{H~{\sc i}}}=11.6$~M$_\odot$ is likely due to an increasingly large mass of baryons in the X-ray phase in hotter, more X-ray luminous groups. Indeed, these groups have line-of-sight velocity dispersions that imply relatively high dynamical masses, and therefore a substantial fraction of baryons may be in the hot phase. The three low-mass CGs with detectable X-ray emission are HCGs~31, 51, and 59, two of which have X-ray emission associated with vigorous star formation \citep{desjardins13}. Excepting HCG~51, these results support our hypothesis that low-mass groups do not heat their IGM to high temperatures. We caution the reader that the hot gas in low-mass systems may have very high entropy, and therefore low density. From the proportionality $L_X\propto\rho^2$, we expect that such gas would be very difficult to detect. Therefore, HCG~51 may in fact have a lower total mass than the others, or it may have a higher fraction of its mass in the hot gas phase (i.e.,~the gas may be low entropy as discussed above). Finally, we note that there are two groups, HCGs~7 and 40, with high total stellar and \hi\ masses but no detectable X-ray emission.

We note that previous work has examined loose groups of galaxies without detectable X-ray emission, and the consensus is that the gas in these groups is too cool to produce significant X-ray luminosity \citep{mulchaey96,rasmussen06}. The lack of hot gas has been explained in two different ways: \citet{mulchaey96} argue that some groups are too low-mass to effectively heat their gas to X-ray temperatures; while \citet{rasmussen06} hypothesize that X-ray underluminous groups are dynamically young (i.e., in the process of collapsing) and have not had sufficient time to virialize their IGM. These ideas are not mutually exclusive, and indeed both may affect the formation of group X-ray halos. Further, we hypothesize that the absence of an observable hot IGM in low-mass groups may be caused by the hot gas having very low density, in which case it would be have very low surface brightness and be nearly impossible to detect, or that the gas is never heated to very high temperatures and instead cools efficiently. Note that our two hypotheses are independent from one another as the gas density must be high for efficient cooling to occur. Unfortunately, our data are insufficient to test the low-density gas and efficient cooling scenarios.

Regarding the explanations offered by \citet{mulchaey96} and \citet{rasmussen06}, we present the two examples of HCGs~31 and 7. In HCG~31, we find an unusually high baryon fraction of $f_b\approx0.36$ (assuming negligible mass in the form of hot gas) due to the high \hi\ mass, while the ratio of stellar mass to dynamical mass is relatively large, and therefore may indicate that the dynamical mass is underestimated. We remind the reader that the dynamical mass is very uncertain due to the small number of galaxies available with which to measure the velocity dispersion, and the magnitude of this uncertainty is unclear. Assuming that most of the baryonic mass in HCG~31 is in the stellar and \hi\ components, then a reasonable baryon fraction implies it is a relatively low-mass system. Thus, we expect HCG~31 to have a very cool virial temperature. Conversely, HCG~7 is a relatively massive group ($M_{\text{dyn}}=10^{12.1}$~M$_\odot$; \citealt{desjardins13}), and therefore we expect a hotter virial temperature based solely upon the total mass. Again, we regard the dynamical mass with caution, though we do note that its stellar mass implies that it is at least an order of magnitude more massive than HCG~31. Thus, HCG~7 seems X-ray underluminous for its mass, though it is not dynamically young. Indeed, \citet{konstantopoulos10} find that the galaxies in HCG~7 show evidence for long-term, enhanced evolution in the group environment without direct, strong interactions. It is unclear how HCG~7 may form an X-ray halo at some later evolutionary stage, if it will at all.

\begin{figure*}
\centering
\includegraphics[width=\textwidth]{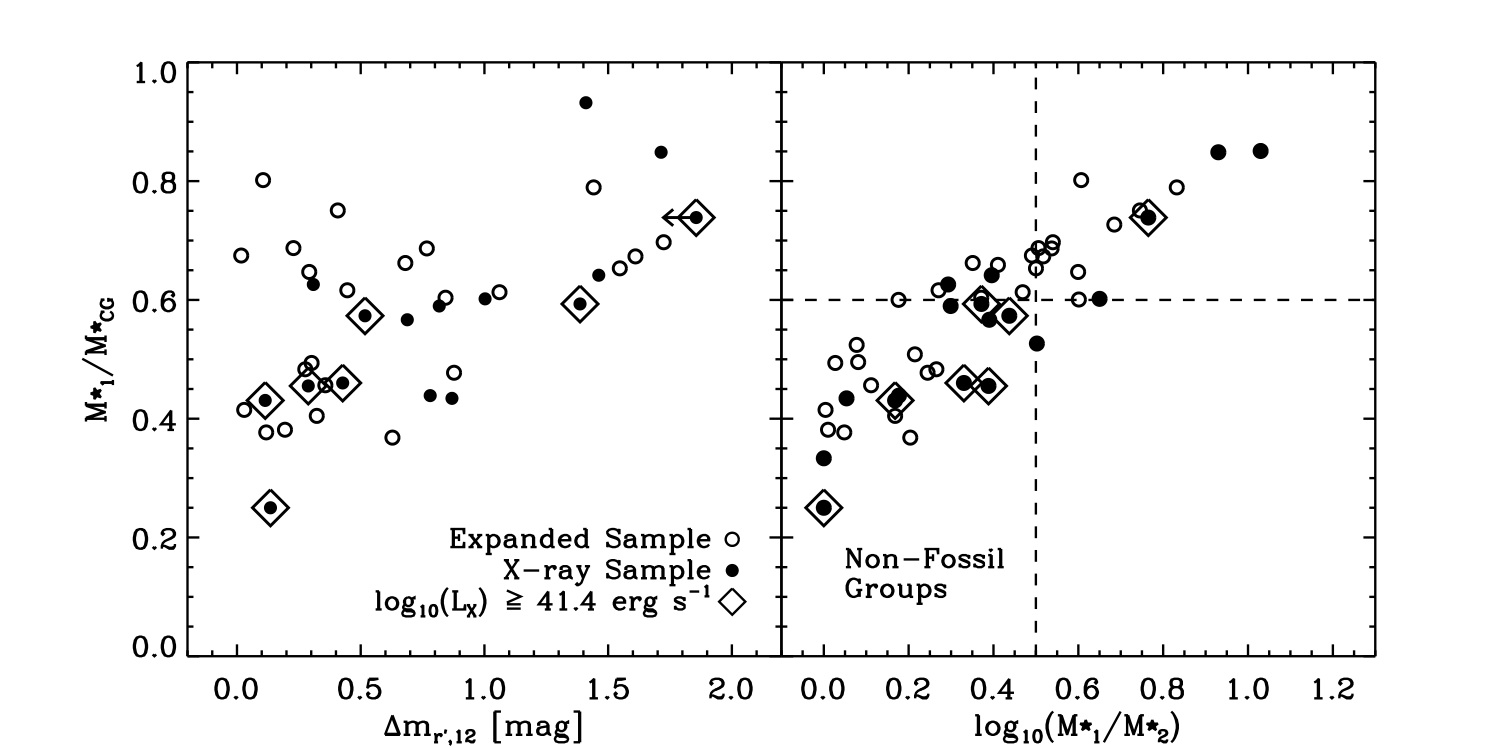}
\caption{The fraction of the total CG stellar mass contained within the most massive group member as a function of the SDSS $r^\prime$ magnitude difference between the two brightest CG galaxies ($\Delta m_{r^\prime,12}$; {\em left}) and the stellar mass ratio between the two most massive group galaxies ({\em right}). Groups that have an X-ray luminosity $\log_{10}(L_X)\geq41.4$~erg~s$^{-1}$ (from the \citealt{jones03} fossil group criteria corrected to our cosmology) are marked with diamonds. One of the magnitude differences in the left panel is an upper-limit because the APO images of HCG~42A and 42B were taken at two difference airmasses. The $r$-band magnitude difference appears to be weakly correlated to the ratio of stellar mass contained within the most massive group galaxy compared to the rest of the group. Therefore, we suggest that the difference in stellar mass between the two most massive galaxies is a better indication of the evolutionary state of the group. The magnitude difference will be related to the stellar mass ratio assuming a single M/L ratio, but this assumption is not justified given the diversity of galaxies found in the most massive CG galaxies.\label{fig:fossil_criteria}}
\end{figure*}

\citet{jones03} observationally define fossil groups to identify groups near the end of their evolution in which the groups are dominated by a single galaxy and have X-ray emission in excess of that associated with normal galaxies. The authors use as their criteria a diffuse X-ray luminosity above a specific threshold ($L_X>10^{42}$~$h_{50}^{-2}$~erg~s$^{-1}$) and a difference in $R$-band magnitude between the two brightest group members $\Delta m_{12}>2$~mag; however, the $\Delta m_{12}$ criterion assumes that the group luminosity function is a well-sampled Schechter function \citep{schechter76}, i.e.,~that the brightest group galaxy is much more luminous than $L^*$, and therefore is located at the bright end of the exponential portion of the luminosity distribution. The \citet{hickson82} CG selection criteria require that group members have $\Delta m\leq3$~mag with respect to the brightest group galaxy, making such a luminosity distribution unlikely, and therefore the \citet{jones03} criteria would have included no HCGs as fossil groups. We note that \citet{dariush10} use a magnitude difference of $\Delta m_{14}>2$~mag, thus identifying more robustly low-mass fossil systems and potentially several CGs, however the optical magnitude selection criterion of the HCG catalog makes it likely that this definition would still miss several fossil groups. Indeed, only HCG~42 comes close to satisfying the \citet{jones03} fossil group selection criteria. We thus seek to form an alternate physically motivated definition of fossil groups in order to classify the CGs in our X-ray sample.

In Figure~\ref{fig:fossil_criteria}, we compare the optical selection criterion from \citet{jones03} against the distribution of stellar mass within the CGs in the \citet{walker12} Expanded Sample\footnote{We excluded RSCGs~67 and 68 from this analysis as they are comprised of galaxies in the core of the Coma Cluster, as well as RSCG~32, which is embedded within the Abell 779 cluster.}. We find that the optical magnitude difference $\Delta m_{12}$ between the brightest galaxies does not trace well the concentration of stellar mass in the CGs, and instead suggest that the fraction of the group stellar mass contained within the first-ranked galaxy (i.e.,~the group member with the largest stellar mass) can be better determined using the difference in stellar mass between the two most massive group galaxies. Note that we strictly use only the two most massive galaxies in the stellar mass criterion, and only the two brightest galaxies in the SDSS $r^\prime$ filter for calculating $\Delta m_{12}$. In groups with two dominant galaxies approximately equal in mass, the most massive and brightest galaxy may not be the same; though, this is only possible when both $\Delta m_{12}$ and $\log_{10}(M_1/M_2)$ are small.

The existence of a lone, massive, early-type galaxy has three possible implications for the groups: that at least one major merger has resulted in the formation of an elliptical galaxy and that only minor mergers will occur in the future; a series of minor mergers has concentrated the bulk of the stellar mass into a cold-gas-poor lenticular galaxy; or there was only ever one massive galaxy in the group and it is now cold-gas-poor. In all three cases, the bulk of the stellar mass exists in a ``red and dead'' galaxy that has reached the end of its evolution in most respects. We therefore use the stellar mass distribution in our classification of CGs as evolved fossil systems. Specifically, we require that: the first-ranked galaxy contain $\geq60\%$ of the group stellar mass; or, alternatively, the first-ranked galaxy be at least a factor of 3 more massive than any other group member. We find that groups evolve from the lower-left region of the right panel of Figure~\ref{fig:fossil_criteria}, in which the two most massive galaxies are of approximately equal mass, to the upper-right region, where groups are dominated by a single massive galaxy. The values used in our fossil group selection criteria (i.e., 60\% of the group stellar mass or a factor of three difference in the stellar masses of the first- and second-ranked galaxies) reflect the distribution of CGs in the right panel of Figure~\ref{fig:fossil_criteria}. Rather than require both of these criteria to be true, we only require that one be satisfied to be considered a fossil group candidate due to the scatter in the CG distribution; in either case the stellar mass is clearly concentrated in a single group member. To ensure that groups we classify as fossil groups are indeed highly evolved, we also impose a morphology criterion requiring that the first-ranked galaxy be a E/S0 galaxy.

We further remove the X-ray criterion from \citet{jones03} because it assumes that the potential of the group is sufficient to virialize the gas to hot temperatures, which is not true in low-mass groups. Indeed, if the gas in low-mass groups is heated to $T\lesssim10^6$~K, then it will cool quickly and may therefore be available again for star formation \citep{dalgarno72,sutherland93,schure09}. We note that the X-ray luminosity requirement from \citet{jones03} was intended to ensure that only truly bound systems were classified as fossil groups. Regardless of if the systems are bound or not, the location of the galaxies in mid-IR color-color space shows evidence for accelerated evolution not observed in other environments \citep{johnson07,walker10,walker12}, therefore the galaxies are in close physical proximity for long enough to affect their evolution in measurable ways. Furthermore, the isolation criterion imposed on the HCGs by \citet{hickson82} makes it unlikely that there are other galaxies to which the CG members may be bound.

Finally, 21 CGs in the Expanded Sample from \citet{walker12} meet our first two fossil group criteria concerning the distribution of stellar mass; however, only six of these CGs (HCGs~19, 22, 40, and 42 and RSCGs 44 and 86) have an E/S0 first-ranked galaxy and are therefore fossil groups under our new definition. From the three fossil CGs with {\em Chandra} observations, only HCG~42 has a substantial X-ray halo according to the \citet{jones03} definition, which requires a clarification of previous findings that fossil groups agree with the galaxy cluster X-ray scaling relations. Our findings suggest that it is only the massive, X-ray luminous fossil groups that agree well with the cluster scaling relations, while low-mass fossil groups systematically fall below these relationships. Furthermore, there may be many fossil groups that have not been classified as such because they are not massive enough to host X-ray luminous halos. Thus, if the X-ray luminous fossil groups represent the origin of optically and X-ray bright, isolated field ellipticals (e.g.,~\citealt{mulchaey99}), then fossil groups with no X-ray emission may be the progenitors of low-mass isolated ellipticals. Indeed, \citet{mulchaey10} find that low-mass field ellipticals are typically X-ray underluminous and suggest that such galaxies, in contrast to their massive counterparts, may not be able to retain hot gas halos primarily due to strong winds from supernovae and, as a secondary factor, AGN. In contrast, \citet{osullivan04} argue that while the galaxy mass may be important in retaining X-ray halos, they hypothesize that these galaxies must also be the result of recent mergers to have sufficiently strong supernovae winds that can drive away low-density hot gas; though the authors also find that the metallicities of field ellipticals do not support supernovae-driven wind models.

\subsection{X-ray Emission and the H~{\sc i} Reservoir}
\label{sec:xrayhi}

\begin{figure}[t]
\centering
\includegraphics[width=\columnwidth]{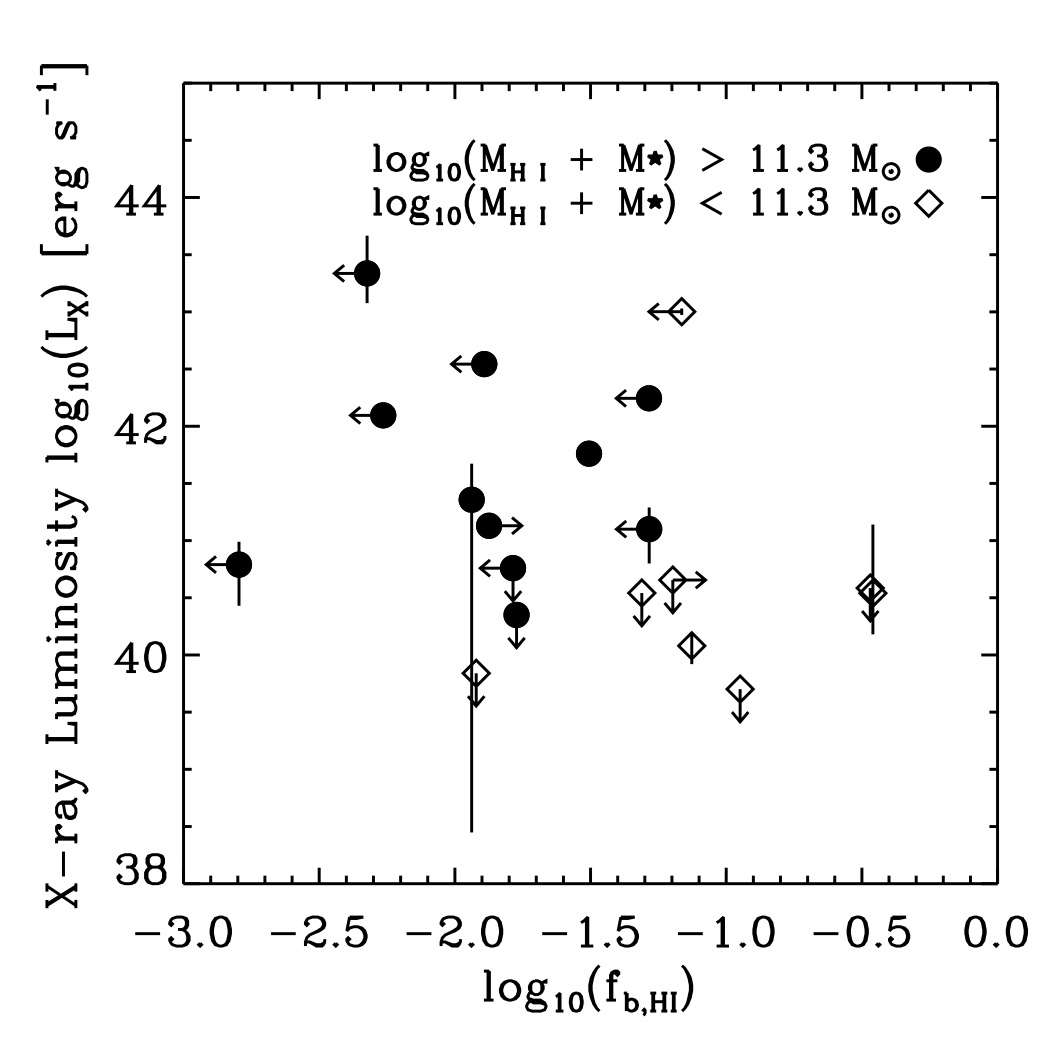}
\caption{The diffuse X-ray luminosity as a function of the \hi\ mass relative to the total stellar and \hi\ mass ($f_{b,\text{H~{\sc i}}}$). Filled circles represent CGs with a total stellar and \hi\ mass of $>10^{11.3}$~M$_\odot$, while empty diamonds are CGs with a total stellar and \hi\ mass of $<10^{11.3}$~M$_\odot$. The lower-limits on the values of $f_{b,\text{H~{\sc i}}}$ represent the groups with additional members that may cause underestimates in the stellar mass measurements. The relatively massive CGs (i.e., those shown as large cirlces) are more luminous in X-rays and extend to smaller values of $f_{b,\text{H~{\sc i}}}$, and therefore more gas in these systems is in the hot rather than cold phase. This is consistent with the most X-ray luminous CGs being more mature, evolved systems.\label{fig:lxhi}}
\end{figure}

In \citet{desjardins13}, the authors compare the X-ray luminosity to the \hi\ mass normalized to the dynamical mass and find tentative evidence to support the hypothesis by \citet{konstantopoulos10} that the X-ray emission in CGs may be dependent upon the morphology of the \hi\ gas in the system. Systems in which the \hi\ has been stripped out of the galaxy disks into the IGM have material to fuel an X-ray luminous halo. We note that the \hi\ stripped from the galaxy disks may not be the only source of fuel for the hot IGM, however a full analysis of the origin of such gas is beyond the scope of this paper. The explanation regarding the morphology of the \hi\ now seems secondary to the mass of the group given our comparison of the X-ray luminosity and the total stellar and \hi\ mass in Section~\ref{sec:baryons}. For example, the mass of the group, and the individual masses of the group members, will cause differences in the distribution of gas in the intragroup space. However, we still compare the X-ray luminosity to the \hi\ reservoir to understand better how neutral gas is consumed in CGs. A concern when using the velocity dispersions, and thus the dynamical masses (as in \citealt{johnson07} and \citealt{desjardins13}), is the inherent uncertainty that stems from the small population of galaxies in each CG. Therefore, to mitigate this uncertainty, we compared the diffuse X-ray luminosity to the ratio of the \hi\ mass to the total stellar and \hi\ mass ($f_{b,\text{H~{\sc i}}}$) in Figure~\ref{fig:lxhi}.

The relatively massive CGs (i.e., total stellar and \hi\ mass $>10^{11}$~M${_\odot}$) are typically more \hi-poor with $0.2\%<f_{b,\text{H~{\sc i}}}<6.6\%$, while the low-mass groups have $5.0\%<f_{b,\text{H~{\sc i}}}<40.6\%$. The most \hi-rich CG is HCG~30 with 40.7\% of the baryonic mass in neutral hydrogen, while the most \hi\ poor CG is HCG~90 with $<0.2$\% of the baryons in neutral gas, where the upper-limit is due to the non-negligible mass of hot gas that we have not included here. As the more massive CGs tend to be more X-ray luminous, it is possible the neutral gas in these systems was virialized to form the X-ray halo before the baryons could be used in star formation. 

\subsection{Comparison of Diffuse X-rays with Star Formation}
\label{subsec:twosfr}

\begin{figure}
\centering
\includegraphics[width=\columnwidth]{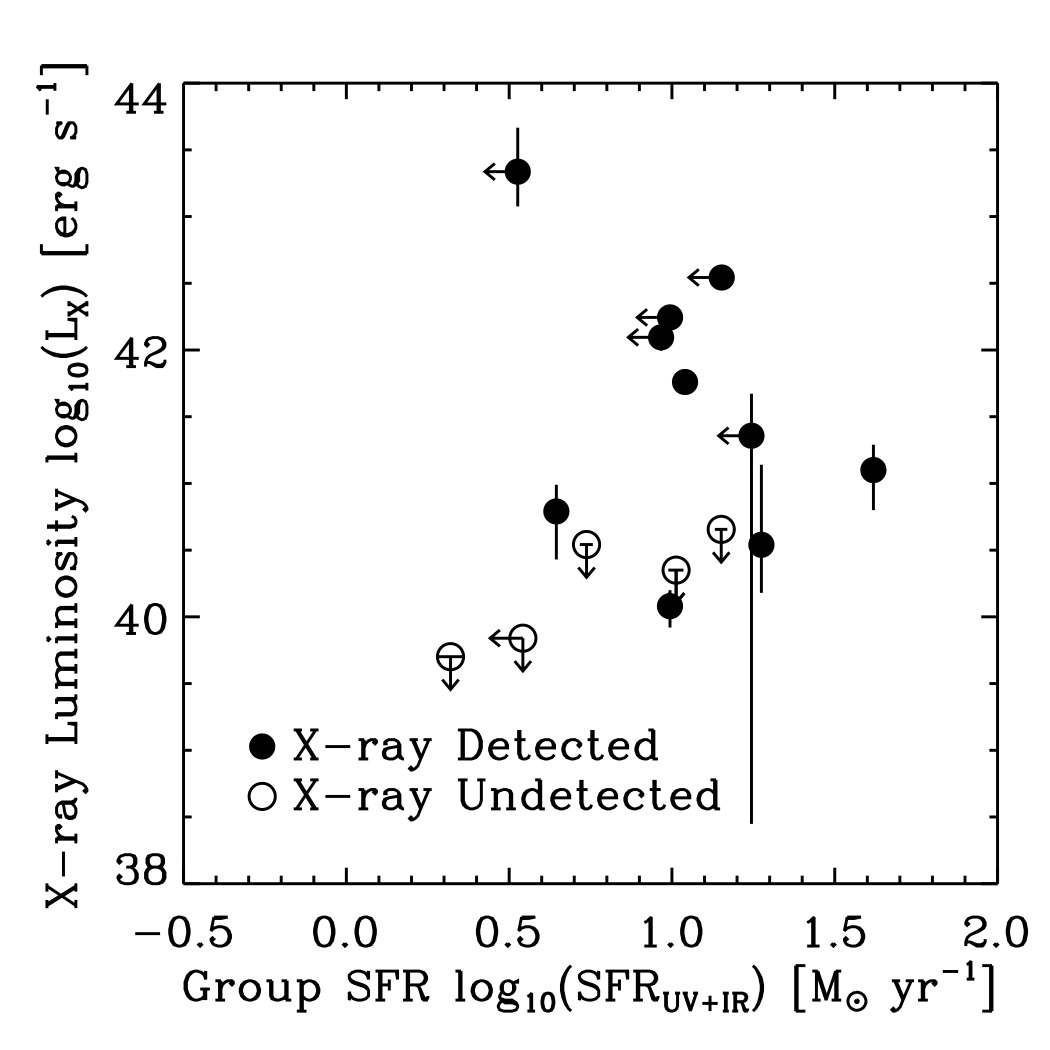}
\caption{The CG diffuse X-ray luminosity as a function of the total group star formation rate as measured from the UV and 24~\micron\ luminosities (Lenki\'c~et~al.~2014,~in~preparation). There are no SFR data for HCGs~30, 40, and 68, and we do not include UV SFRs for HCGs~100C and 100D as these data were not available. An upper-limit on the total SFR occurs when none of the galaxies in a group has detectable star formation. Groups with $L_X\lesssim10^{41}$~erg~s$^{-1}$ (HCGs~16, 31, 59, and 90) tend to have galaxy-linked X-ray emission that increases with total group SFR, while the more X-ray luminous CGs have much lower total SFRs for their X-ray luminosity.\label{fig:lxsfr}}
\end{figure}

\citet{desjardins13} find that CGs with low specific star formation rates (sSFRs; i.e., the star formation rates normalized to the galaxy stellar masses) are X-ray bright compared to CGs with higher sSFRs. This separates X-ray detected CGs into two types: (1) those with gas temperatures and luminosities consistent with virialization; and (2) those with hot gas associated with vigorous star formation ($\text{sSFR}>1.5\times10^{-11}$~yr$^{-1}$). This latter class of star-forming groups appears to be a distinct class in $L_X-\text{sSFR}$ space. Diffuse emission is not detected in HCGs~7 and 22, which are groups with intermediate sSFRs.

Using the UV+24~\micron\ star formation rates (SFRs) from Lenki\'c et al.~({\em in prep.}), which expands upon previous work by \citet{tzanavaris10}, for the \citet{walker12} Expanded Sample of CGs, Figure~\ref{fig:lxsfr} shows the diffuse X-ray luminosity as a function of the total group SFR. Note that SFR data are not available for HCGs~30, 40, and 68. Furthermore, the UV photometry of all of the galaxies in HCG~62 as well as galaxies HCGs~100C and D was unavailable, therefore we consider only IR SFR measurements for these galaxies. The purpose of combining the UV and 24~\micron\ SFRs is to ensure a complete census of star formation over the past $\sim$100~Myr in both areas with low dust column densities and star forming regions enshrouded in dust. The lack of either a UV or IR SFR for a single galaxy is not expected to affect the SFR of that galaxy by a factor of $\gtrsim2$ for normal galaxies (excluding, e.g., LIRGs). We compare the diffuse X-ray emission against the SFRs rather than the sSFRs because it is the absolute rate of star formation that is most likely responsible for the X-ray emission when it is linked to individual galaxies. 


We find that the CGs in the $L_X-\text{SFR}$ space show two distinct distributions. At $L_X\lesssim10^{41}$~erg~s$^{-1}$ (HCGs~16, 31, 59, and 90), the hot gas is galaxy-linked, and the $L_X-\text{SFR}$ relation is found to be generally positive. This is expected if star formation is the source of the hot gas --- more vigorous star formation leads to increased gas heating. Such heating may occur through various processes such as intense ionizing radiation from massive OB associations or supernova shocks. Further study of more groups with high total SFRs and low X-ray luminosities is required to better quantify this result. We note that the X-ray emission in HCG~90 is more likely due to tidal heating rather than star formation \citep{desjardins13}. For CGs with $L_X\gtrsim10^{41}$~erg~s$^{-1}$, the SFR is generally lower in comparison. This overall decrease in SFR for X-ray bright systems may be attributed to gas stripping caused by the hot IGM (i.e., ram-pressure stripping; \citealt{gunn72}), exhaustion of galaxy gas supplies to produce the hot IGM, or the cool gas supply was exhausted by star formation and is not necessarily coupled to the intragroup X-ray emitting gas. 

To investigate the star formation history, and how that relates to the X-ray luminosity of the groups, we also examine the $g^\prime-r^\prime$, $g^\prime-i^\prime$, and $r^\prime-i^\prime$ optical colors of the first-ranked galaxy using data from \citet{walker13}. We note that HCGs~30, 42, 62, 68, and 90 are missing SDSS photometry. From the CGs with SDSS coverage, there are no obvious distinctions in color-color space between the first-ranked galaxies hosted in either X-ray luminous or X-ray non-detected CGs. There were also no correlations between the X-ray luminosity and any of the optical colors themselves. There is no evidence that the X-ray halos of the non-star-forming but X-ray luminous CGs were built-up by a recent ($\lesssim1$~Gyr) burst of star-formation in the first-ranked galaxy.

\section{Summary and Conclusions}
\label{sec:twoconclude}

From a sample of ten CGs, we detect five in X-rays using {\em Chandra} ACIS observations. We combine the CG X-ray temperatures and luminosities with those from \citet{desjardins13} to create a larger sample to examine the buildup of hot gas in CGs. The X-ray detected CGs in the combined sample range from $0.29-1.36$~keV and $10^{40.08-42.73}$~erg~s$^{-1}$ in temperature and X-ray luminosity, respectively. We then compared the X-ray properties against the stellar and \hi\ masses, line-of-sight velocity dispersions, and star formation rates. Our results can be summarized as:

\begin{enumerate}

\item From the cluster scaling laws (Section~\ref{subsec:scaling}; Figure~\ref{fig:lxscale}), we confirm our previous finding in \citet{desjardins13} that cool CGs with low velocity dispersions do not have cluster-like diffuse X-ray emission. However, we now find evidence that relatively massive and X-ray luminous CGs represent a population of groups that are consistent with the X-ray cluster scaling relations. This is consistent with previous X-ray studies of groups such as \citet{connelly12} and \citet{lovisari13}.

\item In Section~\ref{sec:baryons}, we create a physically motivated definition of fossil groups based on the stellar mass distribution and the morphology of the most massive group galaxy, with no additional requirement for an X-ray bright halo. The stellar mass requirement follows from the description of a fossil group as an evolved system in which most of the mass is concentrated in one galaxy, which is supported by the work of \citet{harrison12}. We require that: ({\em a}) the most massive galaxy contain $>60\%$ of the total group stellar mass; or, alternatively, the most massive galaxy contain $\gtrsim3$ times more stellar mass than the next most massive galaxy; and ({\em b}) the most massive galaxy have an E/S0 morphology. This definition is effective at identifying low-mass fossil systems that may have been excluded by previous definitions (e.g., \citealt{jones03}). 

\item Using our fossil group definition, we identify 21 CGs that meet the stellar mass distribution criteria (Figure~\ref{fig:fossil_criteria}), but only six of these (HCGs~19, 22, 40, and 42, and RSCGs~44, and 86) host E/S0 galaxies as the most massive member, and are therefore fossil groups. The high fraction of groups (46\%) with more than 60\% of their stellar mass concentrated in the first-ranked galaxy is similar to the results of \citet{connelly12} for optically and X-ray selected groups. From the three CGs that have {\em Chandra} observations, only HCG~42 is X-ray luminous, and therefore may merge to form a massive, X-ray bright field elliptical. Conversely, HCGs~22 and 40 may merge to be X-ray faint field ellipticals. Furthermore, the spiral dominated groups that meet the stellar mass criteria for fossil groups (HCGs~2, 4, 25, 26, 47, 56, 71, 79, and 100, and RSCGs~31, 34, 64, and 66)\footnote{This list excludes HCG~31 because the aperture for the mid-IR photometry of 31A includes 31C, and HCG~54 because this system is a false group and is instead made up of knots in a single galaxy.} may represent interesting systems for study because they may be the dynamically young precursors to fossil systems.


\item From Sections~\ref{sec:xrayhi} and \ref{subsec:twosfr}, the X-ray luminosity of the CGs in our sample is correlated with the total stellar and \hi\ mass of the groups (Figure~\ref{fig:lx_baryon}), while the relationship with SFR is more complicated. At low values of $L_X$, we qualitatively observe a positive trend between $L_X$ and SFR, while the opposite is true in X-ray bright systems (see Figure~\ref{fig:lxsfr}). Thus, X-ray luminous groups, especially those consistent with the cluster scaling relations, likely represent a population of mature, evolved systems. Low-mass groups are not a homogenous population, and contain both dynamically young (e.g., HCG~31) and dynamically evolved (e.g., HCG~22) systems. In general, low-mass groups, including those classified as fossil systems, are not X-ray luminous and fall well below the X-ray luminosities predicted by the cluster scaling relations.

\end{enumerate}

This last point raises the question of which mechanism takes precedence in the consumption of neutral gas in massive groups: star formation or IGM virialization. Specifically, when does the X-ray halo form in the evolutionary history of the group? The existence of HCGs~7 and 40, which represent relatively massive CGs with no evident X-ray emission and moderate SFRs, may indicate that the formation of the X-ray IGM occurs at a later evolutionary stage, supporting the hypothesis of \citet{rasmussen06} that X-ray underluminous loose groups are relatively dynamically young. HCG~16, another example of a high-mass group, and one that appears to be dynamically young \citep{konstantopoulos13}, was found to be dominated by galaxy-linked X-ray emission; however, a recent deep Chandra observation shows signs that there may be very faint diffuse X-ray emission in the intragroup space of HCG~16 (J. Vrtilek, private communication), and we refer the reader to the analysis of these data (O'Sullivan et al., {\em in prep}). We note that in the case of HCG~7, the group has a low fraction of its mass in cool gas, therefore it may consume its gas reservoir prior in star formation before it is able to heat it to X-ray temperatures. Additionally, the development of a hot IGM could explain the anti-correlation of $L_X$ with SFR as the retardation in star formation due to ram-pressure stripping of cool gas in the group members. This puts our previous question into another perspective: when does star formation in galaxy groups end? The example of RSCG~31, a group with a low SFR, low \hi\ to stellar mass ratio, and high stellar to dynamical mass fraction, may be representative of a system in which star formation consumed the baryons and prevented the formation of an X-ray halo.

Our findings indicate that potentially many low-mass galaxy groups are not X-ray luminous, and therefore X-ray surveys used to identify groups may create heavily biased samples that miss non-negligible fractions of the baryonic mass contained within groups of galaxies (see, e.g., \citealt{rasmussen06,connelly12}). Indeed, if the baryons are not in the molecular, \hi, or X-ray gas, then the evidence indicates that low-mass groups must contain a substantial fraction of baryons in either stars or a warm $10^5$--$10^6$~K gas phase. This warm gas would cool efficiently and be available again for star formation, but it would likely be replenished by gas heating in the group potential. We note that HCGs~30 and 31 have $\gtrsim30$\% of their mass in \hi, however the next most \hi-rich CG contains only 11.3\% of its baryonic mass in neutral gas, indicating that CGs with such an abundance of \hi\ are exceptional rather than commonplace. Furthermore, it is unlikely that the stellar baryon fraction in low-mass groups is higher than in high-mass systems, as other studies find that the ratio of stellar to dynamical mass in groups is $\sim1$\% \citep{andreon10,balogh11,connelly12}. Thus, X-ray surveys only select the X-ray luminous, and therefore massive, groups while missing a significant population of low-mass groups. 

\acknowledgments{T.D.D. and S.C.G. thank the Natural Science and Engineering Research Council of Canada and the Ontario Early Researcher Award Program for support. W.N.B. is supported by NASA ADP grant NNX10AC99G and NSF grant AST-1108604 for support. J.C.C. thanks NSF for funding under award AST-0908984. This work was partially supported
by the ACIS Instrument Team contract SAO SV4-74018 (PI: G. Garmire).

This research has made use of the NASA/IPAC Extragalactic Database (NED) which is operated by the Jet Propulsion Laboratory, California Institute of Technology, under contract with the National Aeronautics and Space Administration. Based on observations obtained with the Apache Point Observatory 3.5-meter telescope, which is owned and operated by the Astrophysical Research Consortium. Funding for SDSS-III has been provided by the Alfred P. Sloan Foundation, the Participating Institutions, the National Science Foundation, and the U.S. Department of Energy Office of Science. The SDSS-III web site is \url{http://www.sdss3.org/}.}

{\em Facilities: } \facility{CXO}

\bibliographystyle{apj}
\bibliography{manuscript,reviewpaper}

\end{document}